\documentclass[nofootinbib,prd]{revtex4}%
\usepackage{amsmath}
\usepackage{amsfonts}
\usepackage{amssymb}
\usepackage{graphicx}%
\begin{document}
\title{Self sustained traversable wormholes and the equation of state}
\author{Remo Garattini}
\email{Remo.Garattini@unibg.it}
\affiliation{Universit\`{a} degli Studi di Bergamo, Facolt\`{a} di Ingegneria, Viale
Marconi 5, 24044 Dalmine (Bergamo) ITALY.}

\begin{abstract}
We compute the graviton one loop contribution to a classical energy in a
\textit{traversable} wormhole background. The form of the shape function
considered is obtained by the equation of state $p=\omega\rho$. We investigate
the size of the wormhole as a function of the parameter $\omega$. The
investigation is evaluated by means of a variational approach with Gaussian
trial wave functionals. A zeta function regularization is involved to handle
with divergences. A renormalization procedure is introduced and the finite one
loop energy is considered as a \textit{self-consistent} source for the
traversable wormhole.The case of the phantom region is briefly discussed.

\end{abstract}
\maketitle

\section{Introduction}

The discovery that our universe is undergoing an accelerated
expansion\cite{expansion} leads to reexamine the Friedmann-Robertson-Walker
equation%
\begin{equation}
\frac{\ddot{a}}{a}=-\frac{4\pi}{3}\left(  \rho+3p\right)  ,
\end{equation}
to explain why the scale factor obeys $\ddot{a}>0$. Indeed, it is evident from
the previous formula, that a sort of \textit{dark energy} is needed to cause a
negative pressure with equation of state%
\begin{equation}
p=\omega\rho.\label{eos}%
\end{equation}
A value of $\omega<-1/3$ is required for the accelerated expansion, while
$\omega=-1$ corresponds to a cosmological constant. A specific form of dark
energy, denoted \textit{phantom energy} has also been proposed with the
property of having $\omega<-1$. It is interesting to note that the phantom
energy violates the null energy condition, $p+\rho<0$, necessary ingredient to
sustain the traversability of wormholes. A wormhole can be represented by two
asymptotically flat regions joined by a bridge. To exist, it must satisfy the
Einstein field equations: one example is represented by the Schwarzschild
solution. One of the prerogatives of a wormhole is its ability to connect two
distant points in space-time. In this amazing perspective, it is immediate to
recognize the possibility of traveling crossing wormholes as a short-cut in
space and time. Unfortunately, although there is no direct evidence, a
Schwarzschild wormhole does not possess this property. It is for this reason
that in a pioneering work Morris and Thorne\cite{MT} and subsequently Morris,
Thorne and Yurtsever\cite{MTY} studied a class of wormholes termed
\textquotedblleft\textit{traversable}\textquotedblright. Unfortunately, the
traversability is accompanied by unavoidable violations of null energy
conditions, namely, the matter threading the wormhole's throat has to be
\textquotedblleft\textit{exotic}\textquotedblright. It is clear that the
existence of dark and phantom energy supports the class of exotic matter. In
this direction, Lobo\cite{Lobo}, Kuhfittig\cite{Kuhfittig} and
Sushkov\cite{Sushkov} have considered the possibility of sustaining the
wormhole traversability with the help of phantom energy. In a previous work,
we explored the possibility that a wormhole can be sustained by its own
quantum fluctuations\cite{Remo0}. In practice, it is the graviton propagating
on the wormhole background that plays the role of the \textquotedblleft%
\textit{exotic}\textquotedblright\ matter. This has not to appear as a
surprise, because the computation involved, namely the one loop contribution
of the graviton to the total energy, is quite similar to compute the Casimir
energy on a fixed background. It is known that, for different physical
systems, Casimir energy is negative and this is exactly one of the features
that the exotic matter should possess. In particular, we conjectured that
quantum fluctuations can support the traversability as effective source of the
semiclassical Einstein's equations. However in Ref.\cite{Remo0}, we limited
the analysis in the region where the equation of state$\left(  \ref{eos}%
\right)  $ assumes the particular value $\omega=1$. In this paper, we will
consider $\omega\in\left(  0,+\infty\right)  $\cite{RemoFrancisco}, although
the semiclassical approach can be judged suspicious because of the suspected
validity of semiclassical methods\footnote{To this purpose, see also paper of
Hochberg, Popov and Sushkov\cite{HPS} and the paper of Khusnutdinov and
Sushkov\cite{KS}.} at the Planck scale\cite{FOP}. The rest of the paper is
structured as follows, in section \ref{p0} we define the effective Einstein
equations, in section \ref{p1} we introduce the traversable wormhole metric,
in section \ref{p2} we give some of the basic rules to perform the functional
integration and we define the Hamiltonian approximated up to second order, in
section \ref{p3} we study the spectrum of the spin-two operator acting on
transverse traceless tensors, in section \ref{p4} we regularize and
renormalize the one loop energy contribution and we speculate about
self-consistency of the result. We summarize and conclude in section \ref{p5}.

\section{The effective Einstein equations}

\label{p0}We begin with a look at the classical Einstein equations%
\begin{equation}
G_{\mu\nu}=\kappa T_{\mu\nu}, \label{Gmunuc}%
\end{equation}
where $T_{\mu\nu}$ is the stress-energy tensor, $G_{\mu\nu}$ is the Einstein
tensor and $\kappa=8\pi G$. Consider a separation of the metric into a
background part, $\bar{g}_{\mu\nu}$, and a perturbation, $h_{\mu\nu}$,%
\begin{equation}
g_{\mu\nu}=\bar{g}_{\mu\nu}+h_{\mu\nu}.
\end{equation}
The Einstein tensor $G_{\mu\nu}$ can also be divided into a part describing
the curvature due to the background geometry and that due to the perturbation,%
\begin{equation}
G_{\mu\nu}\left(  g_{\alpha\beta}\right)  =G_{\mu\nu}\left(  \bar{g}%
_{\alpha\beta}\right)  +\Delta G_{\mu\nu}\left(  \bar{g}_{\alpha\beta
},h_{\alpha\beta}\right)  ,
\end{equation}
where, in principle $\Delta G_{\mu\nu}\left(  \bar{g}_{\alpha\beta}%
,h_{\alpha\beta}\right)  $ is a perturbation series in terms of $h_{\mu\nu}$.
In the context of semiclassical gravity, Eq.$\left(  \ref{Gmunuc}\right)  $
becomes%
\begin{equation}
G_{\mu\nu}=\kappa\left\langle T_{\mu\nu}\right\rangle ^{ren}, \label{Gmunus}%
\end{equation}
where $\left\langle T_{\mu\nu}\right\rangle ^{ren}$ is the renormalized
expectation value of the stress-energy tensor operator of the quantized field.
If the matter field source is absent, nothing prevents us from defining an
effective stress-energy tensor for the fluctuations as\footnote{Note that our
approach is very close to the gravitational \textit{geon} considered by
Anderson and Brill\cite{AndersonBrill}. The relevant difference is in the
averaging procedure.}%
\begin{equation}
\left\langle T_{\mu\nu}\right\rangle ^{ren}=-\frac{1}{\kappa}\left\langle
\Delta G_{\mu\nu}\left(  \bar{g}_{\alpha\beta},h_{\alpha\beta}\right)
\right\rangle ^{ren}.
\end{equation}
From this point of view, the equation governing quantum fluctuations behaves
as a backreaction equation. If we fix our attention to the energy component of
the Einstein field equations, we need to introduce a time-like unit vector
$u^{\mu}$ such that $u\cdot u=-1$. Then the semi-classical Einstein's
equations $\left(  \ref{Gmunus}\right)  $ projected on the constant time
hypersurface $\Sigma$ become%
\begin{equation}
G_{\mu\nu}\left(  \bar{g}_{\alpha\beta}\right)  u^{\mu}u^{\nu}=\kappa
\left\langle T_{\mu\nu}u^{\mu}u^{\nu}\right\rangle ^{ren}=-\left\langle \Delta
G_{\mu\nu}\left(  \bar{g}_{\alpha\beta},h_{\alpha\beta}\right)  u^{\mu}u^{\nu
}\right\rangle ^{ren}.
\end{equation}
To further proceed, it is convenient to consider the associated tensor density
and integrate over $\Sigma$. This leads to\footnote{Details on sign
conventions and decomposition of the Einstein tensor can be found in Apeendix
\ref{app1}}%
\begin{equation}
\frac{1}{2\kappa}\int_{\Sigma}d^{3}x\sqrt{^{3}\bar{g}}G_{\mu\nu}\left(
\bar{g}_{\alpha\beta}\right)  u^{\mu}u^{\nu}=-\int_{\Sigma}d^{3}%
x\mathcal{H}^{\left(  0\right)  }=-\frac{1}{2\kappa}\int_{\Sigma}d^{3}%
x\sqrt{^{3}\bar{g}}\left\langle \Delta G_{\mu\nu}\left(  \bar{g}_{\alpha\beta
},h_{\alpha\beta}\right)  u^{\mu}u^{\nu}\right\rangle ^{ren}, \label{inteq}%
\end{equation}
where%
\begin{equation}
\mathcal{H}^{\left(  0\right)  }=\frac{2\kappa}{\sqrt{^{3}\bar{g}}}G_{ijkl}%
\pi^{ij}\pi^{kl}-\frac{1}{2\kappa}\sqrt{^{3}\bar{g}}R^{\left(  3\right)  }
\label{hdens}%
\end{equation}
is the background field super-hamiltonian and $G_{ijkl}$ is the DeWitt super
metric. Thus the fluctuations in the Einstein tensor are, in this context, the
fluctuations of the hamiltonian. To compute the expectation value of the
perturbed Einstein tensor in the transverse-traceless sector, we use a
variational procedure with gaussian wave functionals. In practice, the right
hand side of Eq.$\left(  \ref{inteq}\right)  $ will be obtained by expanding%
\begin{equation}
E_{wormhole}=\frac{\left\langle \Psi\left\vert H_{\Sigma}\right\vert
\Psi\right\rangle }{\left\langle \Psi|\Psi\right\rangle }=\frac{\left\langle
\Psi\left\vert H_{\Sigma}^{\left(  0\right)  }+H_{\Sigma}^{\left(  1\right)
}+H_{\Sigma}^{\left(  2\right)  }+\ldots\right\vert \Psi\right\rangle
}{\left\langle \Psi|\Psi\right\rangle }%
\end{equation}
and retaining only quantum fluctuations contributing to the effective stress
energy tensor. $H_{\Sigma}^{\left(  i\right)  }$ represents the hamiltonian
approximated to the $i^{th}$ order in $h_{ij}$ and $\Psi$ is a \textit{trial
wave functional} of the gaussian form. Then Eq.$\left(  \ref{inteq}\right)  $
becomes%
\begin{equation}
H_{\Sigma}^{\left(  0\right)  }=\int_{\Sigma}d^{3}x\mathcal{H}^{\left(
0\right)  }=-\frac{\left\langle \Psi\left\vert H_{\Sigma}^{\left(  1\right)
}+H_{\Sigma}^{\left(  2\right)  }+\ldots\right\vert \Psi\right\rangle
}{\left\langle \Psi|\Psi\right\rangle }. \label{flham}%
\end{equation}
The chosen background to compute the quantity contained in Eq.$\left(
\ref{inteq}\right)  $ will be that of a traversable wormhole.

\section{Einstein field equations and the traversable wormhole metric}

\label{p1}In Schwarzschild coordinates, the traversable wormhole metric can be
cast into the form%
\begin{equation}
ds^{2}=-\exp\left(  -2\phi\left(  r\right)  \right)  dt^{2}+\frac{dr^{2}%
}{1-\frac{b\left(  r\right)  }{r}}+r^{2}\left[  d\theta^{2}+\sin^{2}\theta
d\varphi^{2}\right]  . \label{metric}%
\end{equation}
where $\phi\left(  r\right)  $ is called the redshift function, while
$b\left(  r\right)  $ is called the shape function. Using the Einstein field
equation%
\begin{equation}
G_{\mu\nu}=\kappa T_{\mu\nu},
\end{equation}
in an orthonormal reference frame, we obtain the following set of equations%
\begin{equation}
\rho\left(  r\right)  =\frac{1}{8\pi G}\frac{b^{\prime}}{r^{2}}, \label{rhob}%
\end{equation}%
\begin{equation}
p_{r}\left(  r\right)  =\frac{1}{8\pi G}\left[  \frac{2}{r}\left(
1-\frac{b\left(  r\right)  }{r}\right)  \phi^{\prime}-\frac{b}{r^{3}}\right]
,
\end{equation}%
\begin{equation}
p_{t}\left(  r\right)  =\frac{1}{8\pi G}\left(  1-\frac{b\left(  r\right)
}{r}\right)  \left[  \phi^{\prime\prime}+\phi^{\prime}\left(  \phi^{\prime
}+\frac{1}{r}\right)  \right]  -\frac{b^{\prime}r-b}{2r^{2}}\left(
\phi^{\prime}+\frac{1}{r}\right)  ,
\end{equation}
in which $\rho\left(  r\right)  $ is the energy density, $p_{r}\left(
r\right)  $ is the radial pressure, and $p_{t}\left(  r\right)  $ is the
lateral pressure. Using the conservation of the stress-energy tensor, in the
same orthonormal reference frame, we get%
\begin{equation}
p_{r}^{\prime}=\frac{2}{r}\left(  p_{t}-p_{r}\right)  -\left(  \rho
+p_{r}\right)  \phi^{\prime}.
\end{equation}
The Einstein equations can be rearranged to give%
\begin{equation}
b^{\prime}=8\pi G\rho\left(  r\right)  r^{2},
\end{equation}%
\begin{equation}
\phi^{\prime}=\frac{b+8\pi Gp_{r}r^{3}}{2r^{2}\left(  1-\frac{b\left(
r\right)  }{r}\right)  }. \label{phi}%
\end{equation}
Now, we introduce the equation of state $p_{r}=\omega\rho$, and using
Eq.$\left(  \ref{rhob}\right)  $, Eq.$\left(  \ref{phi}\right)  $ becomes%
\begin{equation}
\phi^{\prime}=\frac{b+\omega b^{\prime}r}{2r^{2}\left(  1-\frac{b\left(
r\right)  }{r}\right)  }.
\end{equation}
The redshift function can be set to a constant with respect to the radial
distance, if%
\begin{equation}
b+\omega b^{\prime}r=0.
\end{equation}
The integration of this simple equation leads to%
\begin{equation}
b\left(  r\right)  =r_{t}\left(  \frac{r_{t}}{r}\right)  ^{\frac{1}{\omega}},
\label{shape}%
\end{equation}
where we have used the condition $b\left(  r_{t}\right)  =r_{t}$. In this
situation, the line element $\left(  \ref{metric}\right)  $ becomes%
\begin{equation}
ds^{2}=-Adt^{2}+\frac{dr^{2}}{1-\left(  \frac{r_{t}}{r}\right)  ^{1+\frac
{1}{\omega}}}+r^{2}\left[  d\theta^{2}+\sin^{2}\theta d\varphi^{2}\right]  ,
\label{line}%
\end{equation}
where $A$ is a constant coming from $\phi^{\prime}=0$ which can be set to one
without loss of generality. The parameter $\omega$ is restricted by the
following conditions%
\begin{equation}
b^{\prime}\left(  r_{t}\right)  <1;\qquad\frac{b\left(  r\right)  }%
{r}\underset{r\rightarrow+\infty}{\rightarrow0}\qquad\Longrightarrow
\qquad\left\{
\begin{array}
[c]{c}%
\omega>0\\
\omega<-1
\end{array}
\right.  .
\end{equation}
Proper radial distance is related to the shape function by%
\begin{equation}
l\left(  r\right)  =\pm\int_{r_{t}}^{r}\frac{dr^{\prime}}{\sqrt{1-\frac
{b_{\pm}\left(  r^{\prime}\right)  }{r^{\prime}}}}=\pm r_{t}\frac{2\omega
}{\omega+1}\sqrt{\rho^{\left(  1+\frac{1}{\omega}\right)  }-1}\,_{2}%
F_{1}\ \left(  \frac{1}{2},\frac{1-\omega}{2\omega+2};\frac{3}{2}%
;1-\rho^{\left(  1+\frac{1}{\omega}\right)  }\right)  , \label{prd}%
\end{equation}
where the plus (minus) sign is related to the upper (lower) part of the
wormhole or universe and where $_{2}F_{1}\ \left(  a,b;c;x\right)  $ is a
hypergeometric function. Two coordinate patches are required, each one
covering the range $\left[  r_{t},+\infty\right)  $. Each patch covers one
universe, and the two patches join at $r_{t}$, the throat of the wormhole
defined by%
\begin{equation}
r_{t}=\min\left\{  r\left(  l\right)  \right\}  . \label{throat}%
\end{equation}
When $\omega=1$, we recover the special case where $b\left(  r\right)
=r_{t}^{2}/r$, then the line element simply becomes%
\begin{equation}
ds^{2}=-dt^{2}+\frac{dr^{2}}{1-\frac{r_{t}^{2}}{r^{2}}}+r^{2}\left[
d\theta^{2}+\sin^{2}\theta d\varphi^{2}\right]  \label{me1}%
\end{equation}
and Eq.$\left(  \ref{prd}\right)  $ simplifies into%
\begin{equation}
l\left(  r\right)  =\pm\int_{r_{t}}^{r}\frac{dr^{\prime}}{\sqrt{1-\frac
{r_{t}^{2}}{r^{\prime2}}}}=\pm\sqrt{r^{2}-r_{t}^{2}}\qquad\Longrightarrow
\qquad r^{2}=l^{2}+r_{t}^{2}.
\end{equation}
The new coordinate $l$ covers the range $-\infty<l<+\infty$. The constant time
hypersurface $\Sigma$ is an Einstein-Rosen bridge with wormhole topology
$S^{2}\times R^{1}$. The Einstein-Rosen bridge defines a bifurcation surface
dividing $\Sigma$ in two parts denoted by $\Sigma_{+}$ and $\Sigma_{-}$. To
concretely compute Eq.$\left(  \ref{flham}\right)  $, we consider on the slice
$\Sigma$ deviations from the wormhole metric of the type
\begin{equation}
g_{ij}=\bar{g}_{ij}+h_{ij}, \label{i3}%
\end{equation}
where $g_{ij}$ is extracted from the line element $\left(  \ref{line}\right)
$ whose form becomes%
\begin{equation}
ds^{2}=-dt^{2}+g_{ij}dx^{i}dx^{j}.
\end{equation}

\section{Energy Density Calculation in Schr\"{o}dinger Representation}

\label{p2}In order to compute the quantity%
\begin{equation}
-\int_{\Sigma}d^{3}x\sqrt{^{3}\bar{g}}\left\langle \Delta G_{\mu\nu}\left(
\bar{g}_{\alpha\beta},h_{\alpha\beta}\right)  u^{\mu}u^{\nu}\right\rangle
^{ren}, \label{deltagmn}%
\end{equation}
we consider the right hand side of Eq.$\left(  \ref{flham}\right)  $. Since
$H_{\Sigma}^{\left(  1\right)  }$ is linear in $h_{ij}$ and $h$, the
corresponding gaussian integral disappears and since%
\begin{equation}
\frac{\sqrt{^{3}g}}{\kappa}G_{\mu\nu}\left(  g_{\alpha\beta}\right)  u^{\mu
}u^{\nu}=-\mathcal{H}, \label{GmunuH}%
\end{equation}
it is clear that the hamiltonian expansion in Eq.$\left(  \ref{flham}\right)
$ does not coincide with the averaged expanded Einstein tensor of Eq.$\left(
\ref{deltagmn}\right)  $ because Eq.$\left(  \ref{GmunuH}\right)  $ involves a
tensor density. Therefore, the correct setting is%
\begin{equation}
\int_{\Sigma}d^{3}x\sqrt{^{3}\bar{g}}\left\langle \Delta G_{\mu\nu}\left(
\bar{g}_{\alpha\beta},h_{\alpha\beta}\right)  u^{\mu}u^{\nu}\right\rangle
^{ren}=\int_{\Sigma}d^{3}x\sqrt{^{3}\bar{g}}\frac{\left\langle \Psi\left\vert
\mathcal{H}^{\left(  2\right)  }-\sqrt{^{3}g}^{\left(  2\right)  }%
\mathcal{H}^{\left(  0\right)  }\right\vert \Psi\right\rangle }{\left\langle
\Psi|\Psi\right\rangle }, \label{GmunuH2}%
\end{equation}
where $\sqrt{^{3}g}^{\left(  2\right)  }$ is the second order expanded tensor
density weight. Following the same procedure of Refs.\cite{Remo,Remo1}, the
potential part of the right hand side of Eq.$\left(  \ref{GmunuH2}\right)  $
becomes%
\begin{equation}
\int_{\Sigma}d^{3}x\sqrt{\bar{g}}\left[  -\frac{1}{4}h\triangle h+\frac{1}%
{4}h^{li}\triangle h_{li}-\frac{1}{2}h^{ij}\nabla_{l}\nabla_{i}h_{j}^{l}%
+\frac{1}{2}h\nabla_{l}\nabla_{i}h^{li}-\frac{1}{2}h^{ij}R_{ia}h_{j}^{a}%
+\frac{1}{2}hR_{ij}h^{ij}+\frac{1}{4}Rh^{ij}h_{ij}-\frac{1}{4}hRh\right]  .
\end{equation}
The term%
\begin{equation}
\int_{\Sigma}d^{3}x\sqrt{\bar{g}}\left[  \frac{1}{4}Rh^{ij}h_{ij}-\frac{1}%
{4}hRh\right]  ,
\end{equation}
makes the difference between the hamiltonian expansion and the Einstein tensor
expansion. To explicitly make calculations, we need an orthogonal
decomposition for both $\pi_{ij\text{ }}$and $h_{ij}$ to disentangle gauge
modes from physical deformations. We define the inner product%

\begin{equation}
\left\langle h,k\right\rangle :=\int_{\Sigma}\sqrt{g}G^{ijkl}h_{ij}\left(
x\right)  k_{kl}\left(  x\right)  d^{3}x,
\end{equation}
by means of the inverse WDW metric $G_{ijkl}$, to have a metric on the space
of deformations, i.e. a quadratic form on the tangent space at h, with%

\begin{equation}%
\begin{array}
[c]{c}%
G^{ijkl}=(g^{ik}g^{jl}+g^{il}g^{jk}-2g^{ij}g^{kl})\text{.}%
\end{array}
\end{equation}
The inverse metric is defined on co-tangent space and it assumes the form%

\begin{equation}
\left\langle p,q\right\rangle :=\int_{\Sigma}\sqrt{g}G_{ijkl}p^{ij}\left(
x\right)  q^{kl}\left(  x\right)  d^{3}x\text{,}%
\end{equation}
so that%

\begin{equation}
G^{ijnm}G_{nmkl}=\frac{1}{2}\left(  \delta_{k}^{i}\delta_{l}^{j}+\delta
_{l}^{i}\delta_{k}^{j}\right)  .
\end{equation}
Note that in this scheme the ``inverse metric'' is actually the WDW metric
defined on phase space. The desired decomposition on the tangent space of
3-metric deformations\cite{BergerEbin,York} is:%

\begin{equation}
h_{ij}=\frac{1}{3}hg_{ij}+\left(  L\xi\right)  _{ij}+h_{ij}^{\bot}
\label{p21a}%
\end{equation}
where the operator $L$ maps $\xi_{i}$ into symmetric tracefree tensors%

\begin{equation}
\left(  L\xi\right)  _{ij}=\nabla_{i}\xi_{j}+\nabla_{j}\xi_{i}-\frac{2}%
{3}g_{ij}\left(  \nabla\cdot\xi\right)  .
\end{equation}
Thus the inner product between three-geometries becomes
\[
\left\langle h,h\right\rangle :=\int_{\Sigma}\sqrt{g}G^{ijkl}h_{ij}\left(
x\right)  h_{kl}\left(  x\right)  d^{3}x=
\]%
\begin{equation}
\int_{\Sigma}\sqrt{g}\left[  -\frac{2}{3}h^{2}+\left(  L\xi\right)
^{ij}\left(  L\xi\right)  _{ij}+h^{ij\bot}h_{ij}^{\bot}\right]  . \label{p21b}%
\end{equation}
With the orthogonal decomposition in hand we can define the trial wave
functional
\begin{equation}
\Psi\left\{  h_{ij}\left(  \overrightarrow{x}\right)  \right\}  =\mathcal{N}%
\exp\left\{  -\frac{1}{4l_{p}^{2}}\left[  \left\langle hK^{-1}h\right\rangle
_{x,y}^{\bot}+\left\langle \left(  L\xi\right)  K^{-1}\left(  L\xi\right)
\right\rangle _{x,y}^{\Vert}+\left\langle hK^{-1}h\right\rangle _{x,y}%
^{Trace}\right]  \right\}  ,
\end{equation}
where $\mathcal{N}$ is a normalization factor. We are interested in
perturbations of the physical degrees of freedom. Thus we fix our attention
only to the TT tensor sector reducing therefore the previous form into
\begin{equation}
\Psi\left\{  h_{ij}\left(  \overrightarrow{x}\right)  \right\}  =\mathcal{N}%
\exp\left\{  -\frac{1}{4}\left\langle hK^{-1}h\right\rangle _{x,y}^{\bot
}\right\}  .
\end{equation}
Therefore to calculate the energy density, we need to know the action of some
basic operators on $\Psi\left[  h_{ij}\right]  $. The action of the operator
$h_{ij}$ on $|\Psi\rangle=\Psi\left[  h_{ij}\right]  $ is realized by
\begin{equation}
h_{ij}\left(  x\right)  |\Psi\rangle=h_{ij}\left(  \overrightarrow{x}\right)
\Psi\left\{  h_{ij}\right\}  .
\end{equation}
The action of the operator $\pi_{ij}$ on $|\Psi\rangle$, in general, is%

\begin{equation}
\pi_{ij}\left(  x\right)  |\Psi\rangle=-i\frac{\delta}{\delta h_{ij}\left(
\overrightarrow{x}\right)  }\Psi\left\{  h_{ij}\right\}  .
\end{equation}
The inner product is defined by the functional integration:
\begin{equation}
\left\langle \Psi_{1}\mid\Psi_{2}\right\rangle =\int\left[  \mathcal{D}%
h_{ij}\right]  \Psi_{1}^{\ast}\left\{  h_{ij}\right\}  \Psi_{2}\left\{
h_{kl}\right\}  ,
\end{equation}
and by applying previous functional integration rules, we obtain the
expression of the one-loop-like Hamiltonian form for TT (traceless and
transverse) deformations
\begin{equation}
H_{\Sigma}^{\bot}=\frac{1}{4}\int_{\Sigma}d^{3}x\sqrt{g}G^{ijkl}\left[
\left(  16\pi G\right)  K^{-1\bot}\left(  x,x\right)  _{ijkl}+\frac{1}{\left(
16\pi G\right)  }\left(  \triangle_{2}\right)  _{j}^{a}K^{\bot}\left(
x,x\right)  _{iakl}\right]  . \label{p22}%
\end{equation}
The propagator $K^{\bot}\left(  x,x\right)  _{iakl}$ comes from a functional
integration and it can be represented as
\begin{equation}
K^{\bot}\left(  \overrightarrow{x},\overrightarrow{y}\right)  _{iakl}:=%
{\displaystyle\sum_{\tau}}
\frac{h_{ia}^{\left(  \tau\right)  \bot}\left(  \overrightarrow{x}\right)
h_{kl}^{\left(  \tau\right)  \bot}\left(  \overrightarrow{y}\right)
}{2\lambda\left(  \tau\right)  },
\end{equation}
where $h_{ia}^{\left(  \tau\right)  \bot}\left(  \overrightarrow{x}\right)  $
are the eigenfunctions of $\triangle_{2}$, whose eigenvalues will be denoted
with $\tilde{E}^{2}\left(  \tau\right)  $. $\tau$ denotes a complete set of
indices and $\lambda\left(  \tau\right)  $ are a set of variational parameters
to be determined by the minimization of Eq.$\left(  \ref{p22}\right)  $. The
expectation value of $H^{\bot}$ is easily obtained by inserting the form of
the propagator into Eq.$\left(  \ref{p22}\right)  $%
\begin{equation}
E\left(  \lambda_{i}\right)  =\frac{1}{4}%
{\displaystyle\sum_{\tau}}
{\displaystyle\sum_{i=1}^{2}}
\left[  \left(  16\pi G\right)  \lambda_{i}\left(  \tau\right)  +\frac
{\tilde{E}_{i}^{2}\left(  \tau\right)  }{\left(  16\pi G\right)  \lambda
_{i}\left(  \tau\right)  }\right]  .
\end{equation}
By minimizing with respect to the variational function $\lambda_{i}\left(
\tau\right)  $, we obtain the total one loop energy for TT tensors%
\begin{equation}
E^{TT}=\frac{1}{4}%
{\displaystyle\sum_{\tau}}
\left[  \sqrt{\tilde{E}_{1}^{2}\left(  \tau\right)  }+\sqrt{\tilde{E}_{2}%
^{2}\left(  \tau\right)  }\right]  . \label{e1loop}%
\end{equation}
The above expression makes sense only for $\tilde{E}_{i}^{2}\left(
\tau\right)  >0$, $i=1,2$. The meaning of $\tilde{E}_{i}^{2}$ will be
clarified in the next section. Coming back to Eq.$\left(  \ref{flham}\right)
$, we observe that the value of the wormhole energy on the chosen background
is\footnote{Details of the calculation can be found in the Appendix
\ref{app}.}%
\begin{equation}
\int_{\Sigma}d^{3}x\mathcal{H}^{\left(  0\right)  }=-\frac{1}{16\pi G}%
\int_{\Sigma}d^{3}x\sqrt{\bar{g}}R^{\left(  3\right)  }=A\left(
\omega\right)  \frac{r_{t}}{G},\qquad\omega>-1 \label{ecl}%
\end{equation}
where%
\begin{equation}
A\left(  \omega\right)  =\frac{1}{1+\omega}B\left(  \frac{1}{2},\frac
{1}{1+\omega}\right)  =\frac{\sqrt{\pi}}{\left(  1+\omega\right)  }%
\frac{\Gamma\left(  \frac{1}{1+\omega}\right)  }{\Gamma\left(  \frac{3+\omega
}{2+2\omega}\right)  }. \label{aomega}%
\end{equation}
$B\left(  x,y\right)  $ is the Beta function and $\Gamma\left(  x\right)  $ is
the gamma function. Then the one loop the self-consistent equation for TT
tensors becomes%
\begin{equation}
A\left(  \omega\right)  \frac{r_{t}}{G}=-E^{TT}. \label{sceq}%
\end{equation}
Note that for the special value of $\omega=1$, we get%
\begin{equation}
\frac{\pi r_{t}}{2G}=-E^{TT},
\end{equation}
in agreement with the result of Ref. \cite{Remo0}. Note also that the
self-consistency on the hamiltonian as a reversed sign with respect to the
energy component of the Einstein field equations. This means that an eventual
stable point for the hamiltonian is an unstable point for the effective energy
momentum tensor and vice versa.

\section{The transverse traceless (TT) spin 2 operator for the traversable
wormhole and the W.K.B. approximation}

\label{p3}In this section, we evaluate the one loop energy expressed by
Eq.$\left(  \ref{e1loop}\right)  $. To this purpose, we begin with the
operator describing gravitons propagating on the background $\left(
\ref{line}\right)  $. The Lichnerowicz operator in this particular metric is
defined by%
\begin{equation}
\left(  \triangle_{2}h^{TT}\right)  _{i}^{j}:=-\left(  \triangle_{T}%
h^{TT}\right)  _{i}^{j}+2\left(  Rh^{TT}\right)  _{i}^{j}-R\left(
h^{TT}\right)  _{i}^{j}, \label{spin2}%
\end{equation}
where the transverse-traceless (TT) tensor for the quantum fluctuation is
obtained by the following decomposition%
\begin{equation}
h_{i}^{j}=h_{i}^{j}-\frac{1}{3}\delta_{i}^{j}h+\frac{1}{3}\delta_{i}%
^{j}h=\left(  h^{T}\right)  _{i}^{j}+\frac{1}{3}\delta_{i}^{j}h.
\end{equation}
This implies that $\left(  h^{T}\right)  _{i}^{i}=0$. The transversality
condition is applied on $\left(  h^{T}\right)  _{i}^{j}$ and becomes
$\nabla_{j}\left(  h^{T}\right)  _{i}^{j}=0$. Thus%
\begin{equation}
-\left(  \triangle_{T}h^{TT}\right)  _{i}^{j}=-\triangle_{S}\left(
h^{TT}\right)  _{i}^{j}+\frac{6}{r^{2}}\left(  1-\frac{b\left(  r\right)  }%
{r}\right)  , \label{tlap}%
\end{equation}
where $\triangle_{S}$ is the scalar curved Laplacian, whose form is%
\begin{equation}
\triangle_{S}=\left(  1-\frac{b\left(  r\right)  }{r}\right)  \frac{d^{2}%
}{dr^{2}}+\left(  \frac{4r-b^{\prime}\left(  r\right)  r-3b\left(  r\right)
}{2r^{2}}\right)  \frac{d}{dr}-\frac{L^{2}}{r^{2}} \label{slap}%
\end{equation}
and $R_{j\text{ }}^{a}$ is the mixed Ricci tensor whose components are:
\begin{equation}
R_{i}^{a}=\left\{  \frac{b^{\prime}\left(  r\right)  }{r^{2}}-\frac{b\left(
r\right)  }{r^{3}},\frac{b^{\prime}\left(  r\right)  }{2r^{2}}+\frac{b\left(
r\right)  }{2r^{3}},\frac{b^{\prime}\left(  r\right)  }{2r^{2}}+\frac{b\left(
r\right)  }{2r^{3}}\right\}  .
\end{equation}
The scalar curvature is%
\begin{equation}
R=R_{i}^{j}\delta_{j}^{i}=2\frac{b^{\prime}\left(  r\right)  }{r^{2}}%
\end{equation}
We are therefore led to study the following eigenvalue equation
\begin{equation}
\left(  \triangle_{2}h^{TT}\right)  _{i}^{j}=\tilde{E}^{2}h_{i}^{j}
\label{p31}%
\end{equation}
where $\omega^{2}$ is the eigenvalue of the corresponding equation. In doing
so, we follow Regge and Wheeler in analyzing the equation as modes of definite
frequency, angular momentum and parity\cite{RW}. In particular, our choice for
the three-dimensional gravitational perturbation is represented by its
even-parity form%

\begin{equation}
h_{ij}^{even}\left(  r,\vartheta,\phi\right)  =diag\left[  H\left(  r\right)
\left(  1-\frac{b\left(  r\right)  }{r}\right)  ^{-1},r^{2}K\left(  r\right)
,r^{2}\sin^{2}\vartheta L\left(  r\right)  \right]  Y_{lm}\left(
\vartheta,\phi\right)  , \label{pert}%
\end{equation}
with%
\begin{equation}
\left\{
\begin{array}
[c]{c}%
H\left(  r\right)  =h_{1}^{1}\left(  r\right)  -\frac{1}{3}h\left(  r\right)
\\
K\left(  r\right)  =h_{2}^{2}\left(  r\right)  -\frac{1}{3}h\left(  r\right)
\\
L\left(  r\right)  =h_{3}^{3}\left(  r\right)  -\frac{1}{3}h\left(  r\right)
\end{array}
\right.  .
\end{equation}
From the transversality condition we obtain $h_{2}^{2}\left(  r\right)
=h_{3}^{3}\left(  r\right)  $. Then $K\left(  r\right)  =L\left(  r\right)  $.
For a generic value of the angular momentum $L$, representation $\left(
\ref{pert}\right)  $ joined to Eq.$\left(  \ref{tlap}\right)  $ lead to the
following system of PDE's%

\begin{equation}
\left\{
\begin{array}
[c]{c}%
\left(  -\triangle_{l}+2\left(  \frac{b^{\prime}\left(  r\right)  }{r^{2}%
}-\frac{b\left(  r\right)  }{r^{3}}-\frac{b^{\prime}\left(  r\right)  }{r^{2}%
}\right)  \right)  H\left(  r\right)  =\tilde{E}_{1,l}^{2}H\left(  r\right) \\
\\
\left(  -\triangle_{l}+2\left(  \frac{b^{\prime}\left(  r\right)  }{2r^{2}%
}+\frac{b\left(  r\right)  }{2r^{3}}-\frac{b^{\prime}\left(  r\right)  }%
{r^{2}}\right)  \right)  K\left(  r\right)  =\tilde{E}_{2,l}^{2}K\left(
r\right)
\end{array}
\right.  , \label{p33}%
\end{equation}
where $\triangle_{l}$ is%

\begin{equation}
\triangle_{l}=\left(  1-\frac{b\left(  r\right)  }{r}\right)  \frac{d^{2}%
}{dr^{2}}+\left(  \frac{4r-b^{\prime}\left(  r\right)  r-3b\left(  r\right)
}{2r^{2}}\right)  \frac{d}{dr}-\frac{l\left(  l+1\right)  }{r^{2}}-\frac
{6}{r^{2}}\left(  1-\frac{b\left(  r\right)  }{r}\right)  . \label{p33a}%
\end{equation}
Defining reduced fields%

\begin{equation}
H\left(  r\right)  =\frac{f_{1}\left(  r\right)  }{r};\qquad K\left(
r\right)  =\frac{f_{2}\left(  r\right)  }{r},
\end{equation}
and passing to the proper geodesic distance from the \textit{throat} of the
bridge defined by%
\begin{equation}
dx=\pm\frac{dr}{\sqrt{1-\frac{b\left(  r\right)  }{r}}},
\end{equation}
the system $\left(  \ref{p33}\right)  $ becomes $\left(  r\equiv r\left(
x\right)  \right)  $%

\begin{equation}
\left\{
\begin{array}
[c]{c}%
\left[  -\frac{d^{2}}{dx^{2}}+V_{1}\left(  r\right)  \right]  f_{1}\left(
x\right)  =\tilde{E}_{1,l}^{2}f_{1}\left(  x\right) \\
\\
\left[  -\frac{d^{2}}{dx^{2}}+V_{2}\left(  r\right)  \right]  f_{2}\left(
x\right)  =\tilde{E}_{2,l}^{2}f_{2}\left(  x\right)
\end{array}
\right.  \label{p34}%
\end{equation}
where we have defined $r\equiv r\left(  x\right)  $ and
\begin{equation}
\left\{
\begin{array}
[c]{c}%
V_{1}\left(  r\right)  =\frac{l\left(  l+1\right)  }{r^{2}}+U_{1}\left(
r\right) \\
\\
V_{2}\left(  r\right)  =\frac{l\left(  l+1\right)  }{r^{2}}+U_{2}\left(
r\right)
\end{array}
\right.  ,
\end{equation}
with%
\begin{equation}
\left\{
\begin{array}
[c]{c}%
U_{1}\left(  r\right)  =c_{1}\left(  r\right)  +\left(  \frac{1}{\omega
}-3\right)  c_{2}\left(  r\right)  ,\\
U_{2}\left(  r\right)  =c_{1}\left(  r\right)  +3\left(  \frac{1}{\omega
}+1\right)  c_{2}\left(  r\right)  ,\\
c_{1}\left(  r\right)  =\frac{6}{r^{2}}\left(  1-\left(  \frac{r_{t}}%
{r}\right)  ^{1+\frac{1}{\omega}}\right)  \qquad c_{2}\left(  r\right)
=\frac{1}{2r^{2}}\left(  \frac{r_{t}}{r}\right)  ^{1+\frac{1}{\omega}}%
\end{array}
\right.  . \label{potentials}%
\end{equation}
In order to use the WKB approximation, we define two r-dependent radial wave
numbers $k_{1}\left(  x,l,\tilde{E}_{1,nl}\right)  $ and $k_{2}\left(
x,l,\tilde{E}_{2,nl}\right)  $%
\begin{equation}
\left\{
\begin{array}
[c]{c}%
k_{1}^{2}\left(  x,l,\tilde{E}_{1,nl}\right)  =\tilde{E}_{1,nl}^{2}%
-\frac{l\left(  l+1\right)  }{r^{2}}-U_{1}\left(  r\right) \\
\\
k_{2}^{2}\left(  x,l,\tilde{E}_{2,nl}\right)  =\tilde{E}_{2,nl}^{2}%
-\frac{l\left(  l+1\right)  }{r^{2}}-U_{2}\left(  r\right)
\end{array}
\right.  . \label{rwn}%
\end{equation}
The number of modes with frequency less than $\tilde{E}_{i}$, $i=1,2$, is
given approximately by%
\begin{equation}
\tilde{g}\left(  \tilde{E}_{i}\right)  =\int\nu_{i}\left(  l,\tilde{E}%
_{i}\right)  \left(  2l+1\right)  dl,
\end{equation}
where $\nu_{i}\left(  l,\omega_{i}\right)  $, $i=1,2$ is the number of nodes
in the mode with $\left(  l,\omega_{i}\right)  $, such that%
\begin{equation}
\nu_{i}\left(  l,\omega_{i}\right)  =\frac{1}{2\pi}\int_{-\infty}^{+\infty
}dx\sqrt{k_{i}^{2}\left(  x,l,\omega_{i}\right)  }.
\end{equation}
Here it is understood that the integration with respect to $x$ and $l$ is
taken over those values which satisfy $k_{i}^{2}\left(  x,l,\tilde{E}%
_{i}\right)  \geq0,$ $i=1,2$. Thus the total one loop energy for TT tensors is
given by (recall that $r\equiv r\left(  x\right)  $)%
\[
E^{TT}=\frac{1}{4}\sum_{i=1}^{2}\int_{0}^{+\infty}\tilde{E}_{i}\frac
{d\tilde{g}\left(  \tilde{E}_{i}\right)  }{d\tilde{E}_{i}}d\tilde{E}_{i}%
=\sum_{i=1}^{2}\int_{-\infty}^{+\infty}dxr^{2}\left[  \frac{1}{4\pi}%
\int_{\sqrt{U_{i}\left(  r\right)  }}^{+\infty}\tilde{E}_{i}^{2}\sqrt
{\tilde{E}_{i}^{2}-U_{i}\left(  r\right)  }d\tilde{E}_{i}\right]
\]%
\begin{equation}
=\int_{r_{t}}^{+\infty}drr^{2}\left[  \rho_{1}+\rho_{2}\right]  ,
\label{tote1loop}%
\end{equation}
where%
\begin{equation}
\left\{
\begin{array}
[c]{c}%
\rho_{1}=\frac{1}{4\pi}\int_{\sqrt{U_{1}\left(  x\right)  }}^{+\infty}%
\tilde{E}_{1}^{2}\sqrt{\tilde{E}_{1}^{2}-U_{1}\left(  r\right)  }d\tilde
{E}_{1}\\
\\
\rho_{2}=\frac{1}{4\pi}\int_{\sqrt{U_{2}\left(  x\right)  }}^{+\infty}%
\tilde{E}_{2}^{2}\sqrt{\tilde{E}_{2}^{2}-U_{2}\left(  r\right)  }d\tilde
{E}_{2}%
\end{array}
\right.  . \label{edens}%
\end{equation}

\section{One loop energy Regularization and Renormalization}

\label{p4}In this section, we proceed to evaluate the one loop energy. The
method is equivalent to the scattering phase shift method and to the same
method used to compute the entropy in the brick wall model. We use the zeta
function regularization method to compute the energy densities $\rho_{1}$ and
$\rho_{2}$. Note that this procedure is completely equivalent to the
subtraction procedure of the Casimir energy computation where zero point
energy (ZPE) in different backgrounds with the same asymptotic properties is
involved. To this purpose, we introduce the additional mass parameter $\mu$ in
order to restore the correct dimension for the regularized quantities. Such an
arbitrary mass scale emerges unavoidably in any regularization schemes. Then
we have%
\begin{equation}
\rho_{i}\left(  \varepsilon\right)  =\frac{1}{4\pi}\mu^{2\varepsilon}%
\int_{\sqrt{U_{i}\left(  r\right)  }}^{+\infty}d\tilde{E}_{i}\frac{\tilde
{E}_{i}^{2}}{\left(  \tilde{E}_{i}^{2}-U_{i}\left(  r\right)  \right)
^{\varepsilon-\frac{1}{2}}} \label{zeta}%
\end{equation}
If one of the functions $U_{i}\left(  r\right)  $ is negative, then the
integration has to be meant in the range where $\tilde{E}_{i}^{2}+U_{i}\left(
r\right)  \geq0$. In both cases the result of the integration
is\footnote{Details of the calculation can be found in the Appendix
\ref{app2}.}%
\begin{equation}
=-\frac{U_{i}^{2}\left(  r\right)  }{64\pi^{2}}\left[  \frac{1}{\varepsilon
}+\ln\left(  \frac{\mu^{2}}{U_{i}\left(  r\right)  }\right)  +2\ln2-\frac
{1}{2}\right]  , \label{zeta1}%
\end{equation}
where the absolute value has been inserted to take account of the possible
change of sign. Then the total regularized one loop energy is%
\begin{equation}
E^{TT}\left(  r_{t},\varepsilon;\mu\right)  =4\pi\left\{  2\int_{r_{t}%
}^{+\infty}dr\frac{r^{2}}{\sqrt{1-\frac{b\left(  r\right)  }{r}}}\left[
\left(  \rho_{1}\left(  \varepsilon\right)  +\rho_{2}\left(  \varepsilon
\right)  \right)  \right]  \right\}  ,
\end{equation}
where the factor $4\pi$ comes from the angular integration, while the factor
$2$ in front of the integral appears because we have come back to the original
radial coordinate $r$: this means that we have to double the computation
because of the upper and lower universe. To further proceed, it is useful to
define the following coefficients:

\begin{enumerate}
\item
\begin{equation}
a:=2r_{t}\int_{r_{t}}^{+\infty}dr\frac{r^{2}}{\sqrt{1-\frac{b\left(  r\right)
}{r}}}\left[  U_{1}^{2}\left(  r\right)  +U_{2}^{2}\left(  r\right)  \right]
,
\end{equation}
and

\item
\begin{equation}
\tilde{b}:=-2r_{t}\int_{r_{t}}^{+\infty}dr\frac{r^{2}}{\sqrt{1-\frac{b\left(
r\right)  }{r}}}\left[  U_{1}^{2}\left(  r\right)  \ln\left\vert U_{1}\left(
r\right)  \right\vert +U_{2}^{2}\left(  r\right)  \ln\left(  U_{2}\left(
r\right)  \right)  \right]  .
\end{equation}%
\begin{equation}
c:=2r_{t}\int_{r_{t}}^{+\infty}dr\frac{r^{2}}{\sqrt{1-\frac{b\left(  r\right)
}{r}}}\left[  U_{1}^{2}\left(  r\right)  +U_{2}^{2}\left(  r\right)  \right]
\ln\left(  \mu^{2}\right)  =2a\ln\mu.
\end{equation}

\end{enumerate}

Rescale the radial coordinate $r$ so that $\rho=r/r_{t}$, then%
\begin{equation}
\frac{b\left(  r\right)  }{r}=\left(  \frac{r_{t}}{r}\right)  ^{1+\frac
{1}{\omega}}=\left(  \frac{1}{\rho}\right)  ^{1+\frac{1}{\omega}};\quad
U_{1}\left(  r\right)  =U_{1}\left(  \rho\right)  /r_{t}^{2};\quad
U_{2}\left(  r\right)  =U_{2}\left(  \rho\right)  /r_{t}^{2}.
\end{equation}
Thus the coefficient $a$ simply becomes%
\begin{equation}
a:=2\int_{1}^{+\infty}d\rho\frac{\rho^{2}}{\sqrt{1-\left(  \frac{1}{\rho
}\right)  ^{1+\frac{1}{\omega}}}}\left[  U_{1}^{2}\left(  \rho\right)
+U_{2}^{2}\left(  \rho\right)  \right]  ,
\end{equation}
while the coefficient $\tilde{b}$ changes into%
\[
\tilde{b}:=-2\int_{1}^{+\infty}d\rho\frac{\rho^{2}}{\sqrt{1-\left(  \frac
{1}{\rho}\right)  ^{1+\frac{1}{\omega}}}}\left[  U_{1}^{2}\left(  \rho\right)
\ln\left\vert U_{1}\left(  \rho\right)  \right\vert +U_{2}^{2}\left(
\rho\right)  \ln\left(  U_{2}\left(  \rho\right)  \right)  \right]  +2\int
_{1}^{+\infty}d\rho\frac{\rho^{2}}{\sqrt{1-\left(  \frac{1}{\rho}\right)
^{1+\frac{1}{\omega}}}}\left[  U_{1}^{2}\left(  \rho\right)  +U_{2}^{2}\left(
\rho\right)  \right]  \left(  2\ln r_{t}\right)
\]%
\[
=-2\int_{1}^{+\infty}d\rho\frac{\rho^{2}}{\sqrt{1-\left(  \frac{1}{\rho
}\right)  ^{1+\frac{1}{\omega}}}}\left[  U_{1}^{2}\left(  \rho\right)
\ln\left\vert U_{1}\left(  \rho\right)  \right\vert +U_{2}^{2}\left(
\rho\right)  \ln\left(  U_{2}\left(  \rho\right)  \right)  \right]  +2a\ln
r_{t}=b+2a\ln r_{t}.
\]

The result of the integration over the $r$ coordinate leads to the following
expression%
\begin{equation}
E^{TT}\left(  r_{t},\varepsilon;\mu\right)  =\frac{1}{16\pi}\left[  -\frac
{a}{\varepsilon r_{t}}-\frac{b}{r_{t}}-\frac{2a}{r_{t}}\ln\left(  \frac
{\sqrt{8}r_{t}\mu}{\sqrt[4]{e}}\right)  \right]  \label{tote1loopreg}%
\end{equation}
and the self consistent equation $\left(  \ref{sceq}\right)  $ can be written
in the form%
\begin{equation}
A\left(  \omega\right)  \frac{r_{t}}{G}=\frac{1}{16\pi}\left[  \frac
{a}{\varepsilon r_{t}}+\frac{b}{r_{t}}+\frac{2a}{r_{t}}\ln\left(  \frac
{\sqrt{8}r_{t}\mu}{\sqrt[4]{e}}\right)  \right]  .\label{sceqtot}%
\end{equation}
Actually, also the coefficients $a$ and $b$ depend on $\omega$, but at this
level is not relevant. In order to deal with finite quantities, we renormalize
the divergent energy by absorbing the singularity in the classical quantity$.$
In particular, we re-define the bare classical constant $G$%
\begin{equation}
\frac{1}{G}\rightarrow\frac{1}{G_{0}}+\frac{a}{\varepsilon A\left(
\omega\right)  16\pi r_{t}^{2}}.
\end{equation}
Therefore, the remaining finite value for the effective equation $\left(
\ref{sceqtot}\right)  $ reads%
\begin{equation}
A\left(  \omega\right)  \frac{r_{t}}{G_{0}}=\frac{1}{16\pi}\left[  \frac
{b}{r_{t}}+\frac{2a}{r_{t}}\ln\left(  \frac{\sqrt{8}r_{t}\mu}{\sqrt[4]{e}%
}\right)  \right]  .\label{sceqtoteff}%
\end{equation}
This quantity depends on the arbitrary mass scale $\mu.$ It is appropriate to
use the renormalization group equation to eliminate such a dependence. To this
aim, we impose that\cite{Cherednikov}%
\begin{equation}
\mu\frac{d}{d\mu}\left(  \frac{A\left(  \omega\right)  r_{t}}{G_{0}\left(
\mu\right)  }\right)  =\mu\frac{d}{d\mu}\left\{  \frac{1}{16\pi}\left[
\frac{b}{r_{t}}+\frac{2a}{r_{t}}\ln\left(  \frac{\sqrt{8}r_{t}\mu}{\sqrt[4]%
{e}}\right)  \right]  \right\}  ,
\end{equation}
namely%
\begin{equation}
A\left(  \omega\right)  r_{t}\mu\frac{\partial G_{0}^{-1}\left(  \mu\right)
}{\partial\mu}-\frac{a}{8\pi r_{t}}=0.
\end{equation}
Solving it we find that the renormalized constant $G_{0}$ should be treated as
a running one in the sense that it varies provided that the scale $\mu$ is
changing
\begin{equation}
\frac{1}{G_{0}\left(  \mu\right)  }=\frac{1}{G_{0}\left(  \mu_{0}\right)
}+K\ln\left(  \frac{\mu}{\mu_{0}}\right)  \qquad\text{or}\qquad G_{0}\left(
\mu\right)  =\frac{G_{0}\left(  \mu_{0}\right)  }{1+G_{0}\left(  \mu
_{0}\right)  K\ln\left(  \frac{\mu}{\mu_{0}}\right)  },\qquad K=\frac
{a}{A\left(  \omega\right)  8\pi r_{t}^{2}};\label{Gmu}%
\end{equation}
where $\mu_{0}$ is the normalization point. We substitute Eq.$\left(
\ref{Gmu}\right)  $ into Eq.$\left(  \ref{sceqtoteff}\right)  $ to find%
\begin{equation}
\frac{A\left(  \omega\right)  }{G_{0}\left(  \mu_{0}\right)  }=\frac{1}{16\pi
}\left[  \frac{b}{r_{t}^{2}}+\frac{2a}{r_{t}^{2}}\ln\left(  \frac{\sqrt
{8}r_{t}\mu_{0}}{\sqrt[4]{e}}\right)  \right]  ,\label{Nsceqtoteff}%
\end{equation}
where we have divided by $r_{t}$. In order to have only one
solution\footnote{Note that in the paper of Khusnutdinov and Sushkov\cite{KS},
to find only one solution, the minimum of the ground state of the quantized
scalar field has been set equal to the classical energy. In our case, we have
no external fields on a given background. This means that it is not possible
to find a minimum of the one loop gravitons, in analogy with Ref.\cite{KS}.
Moreover the renormalization procedure in Ref.\cite{KS} is completely
independent by the classical term, while in our case it is not. Indeed, thanks
to the self-consistent equation $\left(  \ref{sceq}\right)  $, we can
renormalize the divergent term.}, we find the extremum of the r.h.s. of
Eq.$\left(  \ref{Nsceqtoteff}\right)  $ and we get%
\begin{equation}
\frac{a-b}{2a}=\ln\left(  \frac{\sqrt{8}\bar{r}_{t}\mu_{0}}{\sqrt[4]{e}%
}\right)  \qquad\Longrightarrow\qquad\bar{r}_{t}=\frac{\sqrt[4]{e}}{\sqrt
{8}\mu_{0}}\exp\left(  \frac{a-b}{2a}\right)  \label{rtmin}%
\end{equation}
and%
\begin{equation}
\frac{1}{G_{0}\left(  \mu_{0}\right)  }=\frac{a\mu_{0}^{2}}{2\pi A\left(
\omega\right)  \sqrt{e}}\exp\left(  -\frac{a-b}{a}\right)  .\label{G0mu}%
\end{equation}
With the help of Eqs.$\left(  \ref{rtmin}\right)  $ and $\left(
\ref{G0mu}\right)  $, Eq.$\left(  \ref{Gmu}\right)  $ becomes%
\[
\frac{1}{G_{0}\left(  \mu\right)  }=\frac{1}{G_{0}\left(  \mu_{0}\right)
}+\frac{a}{A\left(  \omega\right)  8\pi\bar{r}_{t}^{2}}\ln\left(  \frac{\mu
}{\mu_{0}}\right)  =\frac{a\mu_{0}^{2}}{A\left(  \omega\right)  2\pi\sqrt{e}%
}\exp\left(  -\frac{a-b}{a}\right)  \left[  1+2\ln\left(  \frac{\mu}{\mu_{0}%
}\right)  \right]
\]%
\begin{equation}
=\frac{1}{G_{0}\left(  \mu_{0}\right)  }\left[  1+2\ln\left(  \frac{\mu}%
{\mu_{0}}\right)  \right]  .
\end{equation}
It is straightforward to see that we have a constraint on $\mu/\mu_{0}$.Indeed
we have to choose%
\begin{equation}
\mu>\mu_{0}\exp\left(  -\frac{1}{2}\right)  =.6065306597\mu_{0},
\end{equation}
otherwise $G_{0}\left(  \mu\right)  $ becomes negative\cite{Remo1}. We have
now two possibilities: 1) we identify $G_{0}\left(  \mu_{0}\right)  $ with the
squared Planck length, then the wormhole radius becomes%
\begin{equation}
\bar{r}_{t}=\sqrt{\frac{a\left(  \omega\right)  }{16\pi A\left(
\omega\right)  }}l_{p},\label{a}%
\end{equation}
where we have reestablished the $\omega$ dependence of the coefficient $a$. It
is useful to write the expression for $\omega\rightarrow+\infty$ and for
$\omega\rightarrow0$. We get%
\begin{equation}
\left\{
\begin{array}
[c]{c}%
\bar{r}_{t}\simeq\left[  \,{\frac{\sqrt{105}}{5\sqrt{\pi}}}\left(  1+\frac
{1}{\omega}\left(  {\frac{449}{420}}-2\,\ln\left(  2\right)  \right)
\,+O\left(  \omega^{-2}\right)  \right)  \right]  l_{p}\qquad\omega
\rightarrow+\infty\\
\\
\bar{r}_{t}\simeq\ \left[  {\frac{\sqrt{30}}{12\sqrt{\pi}\sqrt{\omega}}%
}+O\left(  \omega^{1/2}\right)  \right]  l_{p}\qquad\omega\rightarrow0
\end{array}
\right.  .\label{as}%
\end{equation}
The following plot shows the behavior of $\bar{r}_{t}$ as a function of
$\omega$.%
\begin{figure}
[ptbh]
\begin{center}
\includegraphics[
height=2.9317in,
width=2.9317in
]%
{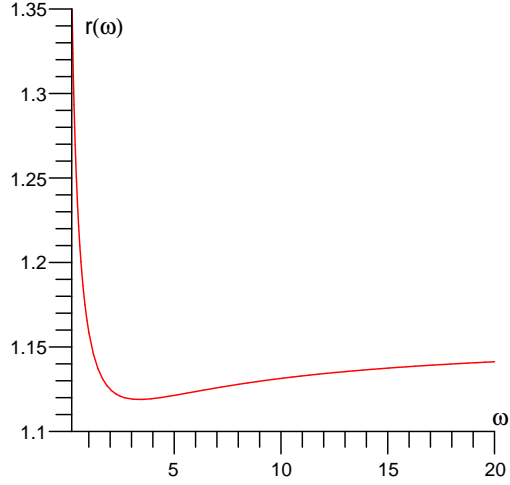}%
\caption{Plot of the wormhole throat $\bar{r}_{t}$ as a function of $\omega$
in the positive range.}%
\label{fig1}%
\end{center}
\end{figure}
It is visible the presence of a minimum for $\bar{\omega}=3.35204$, where
$\bar{r}_{t}\left(  \bar{\omega}\right)  =1.11891$. As we can see, from the
expression $\left(  \ref{as}\right)  $ and from the Fig.\ref{fig1}, the radius
is divergent when $\omega\rightarrow0$. At this stage, we cannot establish if
this is a physical result or a failure of the scheme. When $\omega
\rightarrow\pm\infty$, $\bar{r}_{t}$ approaches the value $\allowbreak
1.15624l_{p}$, while for $\omega=1$, we obtained $\bar{r}_{t}=\allowbreak
1.15882l_{p}$. It is interesting to note that when $\omega\rightarrow+\infty$,
the shape function $b\left(  r\right)  $ in Eq.$\left(  \ref{shape}\right)  $
approaches the Schwarzschild value, when we identify $\bar{r}_{t}$ with $2MG$.
In this sense, it seems that also the Schwarzschild wormhole is
traversable.\pagebreak

2) We identify $\mu_{0}$ with the Planck scale and we get from Eq.$\left(
\ref{rtmin}\right)  $ the following plot%
\begin{figure}
[th]
\begin{center}
\includegraphics[
height=3.442in,
width=3.442in
]%
{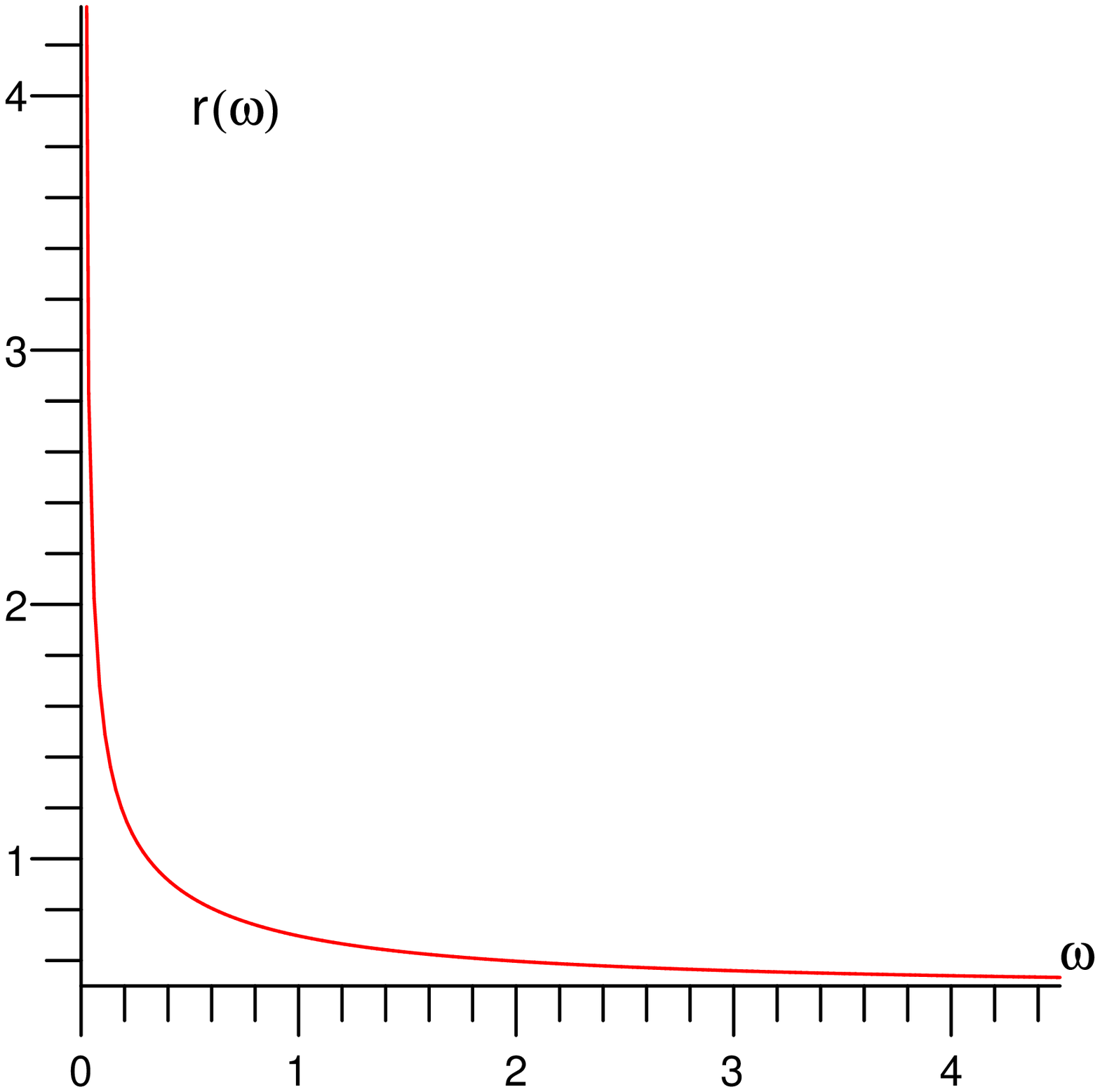}%
\label{f4}%
\end{center}
\end{figure}
Note the absence of a minimum. Differently from case of Eq.$\left(
\ref{a}\right)  $, there is no access to the phantom range. In our case, we
suppose that the graviton quantum fluctuations play the role of the exotic
matter and even if we fix the renormalization point at the Planck scale as in
Ref.\cite{KS}, we find a radius $\bar{r}_{t}>r_{w}$.

\section{Summary and Conclusions}

\label{p5}In this paper, we have generalized the analysis of self-sustaining
wormholes\cite{Remo0} by looking how the equation of state $\left(
\ref{eos}\right)  $ can affect the traversability, when the sign of the
parameter $\omega$ is positive. The paper has been motivated by the work of
Lobo\cite{Lobo}, Kuhfittig\cite{Kuhfittig} and Sushkov\cite{Sushkov} , where
the authors search for classical traversable wormholes supported by phantom
energy. Since the phantom energy must satisfy the equation of state, but in
the range $\omega<-1$, we have investigated the possibility of studying the
whole range $-\infty<\omega<+\infty$. Unfortunately, evaluating the classical
term we have discovered that such a term is well defined in the range
$-1<\omega<+\infty$. The interval $-1<\omega<0$ should be interesting for the
existence of a \textquotedblleft\textit{dark}\textquotedblright\ energy
support. Once again, the \textquotedblleft\textit{dark}\textquotedblright%
\ energy domain lies outside the asymptotically flatness property. So, unless
one is interested in wormholes that are not asymptotically flat, i.e.
asymptotically de Sitter or asymptotically Anti-de Sitter, we have to reject
also this possibility. Therefore, the final stage of computation has been
restricted only to positive values of the parameter $\omega$. In this context,
it is interesting to note that also the Schwarzschild wormhole is traversable,
even if in the limiting procedure of $\omega\rightarrow+\infty$. Despite of
this, the obtained \textquotedblleft\textit{traversability}\textquotedblright%
\ has to be regarded as in \textquotedblleft\textit{principle}%
\textquotedblright\ rather than in \textquotedblleft\textit{practice}%
\textquotedblright\ because the wormhole radius has a Planckian size. We do
not know , at this stage of the calculation, if a different approach for a
self sustained wormhole can give better results. On the other hand, the
positive $\omega$ sector seems to corroborate the Casimir process of the
quantum fluctuations supporting the opening of the wormhole. Even in this
region, we do not know what happens approaching directly the point $\omega=0$,
because it seems that this approach is ill defined. Nevertheless, in this
paper we have studied the behavior of the energy. Work in progress seems to
show that dealing with energy density one can get more general results even in
the \textquotedblleft\textit{phantom}\textquotedblright%
\ sector\cite{RemoFrancisco}.

\acknowledgments

The author would like to thank S. Capozziello, F. Lobo and the Referee for useful
comments and suggestions. In particular, the author would like to thank the Referee,
 for having brought to his attention the paper of Ref.\cite{Sushkov}.
\appendix

\section{Computation of the classical term}

\label{app}Here, we give details leading to Eqs.$\left(  \ref{ecl}%
,\ref{aomega}\right)  $. We begin with the definition of the hamiltonian which
in the static case simplifies into%
\begin{equation}
\int_{\Sigma}d^{3}x\mathcal{H}^{\left(  0\right)  }=-\frac{1}{16\pi G}%
\int_{\Sigma}d^{3}x\sqrt{\bar{g}}R^{\left(  3\right)  }.
\end{equation}
With the help of Eqs$\left(  \ref{shape},\ref{line}\right)  $ , we get%
\begin{equation}
\frac{1}{G\omega}\int_{r_{t}}^{+\infty}dr\left(  \frac{r_{t}}{r}\right)
^{\frac{x}{2}}\frac{1}{\sqrt{\left(  \frac{r}{r_{t}}\right)  ^{x}-1}}\qquad
x:=1+\frac{1}{\omega}.
\end{equation}
In the previous integral there is an extra factor \textquotedblleft%
2\textquotedblright\ coming from the counting of the universes. We change
variable to obtain%
\begin{equation}
\frac{2r_{t}}{G\omega x}\int_{0}^{+\infty}\frac{dt}{\cosh^{2-2/x}\left(
t\right)  }=\frac{r_{t}}{G\left(  1+\omega\right)  }B\left(  \frac{1}{2}%
,\frac{1}{1+\omega}\right)  =\frac{r_{t}}{G}A\left(  \omega\right)
\label{intE}%
\end{equation}
with%
\begin{equation}
A\left(  \omega\right)  =\frac{1}{\left(  1+\omega\right)  }B\left(  \frac
{1}{2},\frac{1}{1+\omega}\right)  =\frac{\sqrt{\pi}}{\left(  1+\omega\right)
}\frac{\Gamma\left(  \frac{1}{1+\omega}\right)  }{\Gamma\left(  \frac
{3+\omega}{2+2\omega}\right)  }.
\end{equation}
In Eq.$\left(  \ref{intE}\right)  $, we have used the following formula%
\begin{equation}
\int_{0}^{+\infty}dt\frac{\sinh^{\mu}t}{\cosh^{\nu}t}=\frac{1}{2}B\left(
\frac{\mu+1}{2},\frac{\nu-\mu}{2}\right)  \qquad\left\{
\begin{array}
[c]{c}%
\operatorname{Re}\mu>-1\\
\operatorname{Re}\left(  \mu-\nu\right)  <0
\end{array}
\right.  .
\end{equation}

\section{Einstein equations and the Hamiltonian}

\label{app1}Let us consider the Einstein equations%
\begin{equation}
G_{\mu\nu}=R_{\mu\nu}-\frac{1}{2}g_{\mu\nu}R=\kappa8\pi GT_{\mu\nu}.
\end{equation}
$R_{\mu\nu}$ is the Ricci tensor and $R$ is the scalar curvature. If $u^{\mu}$
is a time-like unit vector such that $g_{\mu\nu}u^{\mu}u^{\nu}=-1,$then the
Einstein tensor $G_{\mu\nu}$ becomes%
\begin{equation}
G_{\mu\nu}u^{\mu}u^{\nu}=R_{\mu\nu}u^{\mu}u^{\nu}-\frac{1}{2}g_{\mu\nu}u^{\mu
}u^{\nu}R=R_{\mu\nu}u^{\mu}u^{\nu}+\frac{1}{2}R. \label{a1}%
\end{equation}
By means of the Gauss-Codazzi equations\cite{HawEll},
\begin{equation}
R=R^{\left(  3\right)  }\pm2R_{\mu\nu}u^{\mu}u^{\nu}\mp K^{2}\pm K_{\mu\nu
}K^{\mu\nu},
\end{equation}
where $K_{\mu\nu}$ is the extrinsic curvature and $R^{\left(  3\right)  }$ is
the three dimensional scalar curvature. For a time-like vector, we take the
lower sign and Eq.$\left(  \ref{a1}\right)  $ becomes%
\begin{equation}
G_{\mu\nu}u^{\mu}u^{\nu}=\frac{1}{2}\left(  R^{\left(  3\right)  }%
+K^{2}-K_{\mu\nu}K^{\mu\nu}\right)  . \label{a2}%
\end{equation}
If the conjugate momentum is defined by%
\begin{equation}
\pi^{\mu\nu}=\frac{\sqrt{^{\left(  3\right)  }g}}{2\kappa}\left(  Kg^{\mu\nu
}-K^{\mu\nu}\right)  ,
\end{equation}
then%
\begin{equation}
K^{2}-K_{\mu\nu}K^{\mu\nu}=\left(  \frac{2\kappa}{\sqrt{^{\left(  3\right)
}g}}\right)  ^{2}\left(  \frac{\pi^{2}}{2}-\pi^{\mu\nu}\pi_{\mu\nu}\right)
\end{equation}
and%
\begin{equation}
\frac{\sqrt{^{\left(  3\right)  }g}}{2\kappa}G_{\mu\nu}u^{\mu}u^{\nu}%
=\frac{\sqrt{^{\left(  3\right)  }g}}{2\kappa}R^{\left(  3\right)  }%
+\frac{2\kappa}{\sqrt{^{\left(  3\right)  }g}}\left(  \frac{\pi^{2}}{2}%
-\pi^{\mu\nu}\pi_{\mu\nu}\right)  =-\mathcal{H}^{\left(  0\right)  },
\end{equation}
namely Eq.$\left(  \ref{hdens}\right)  $ with the reversed sign.

\section{Energy coefficients}

\label{appe}In this appendix, we report details on computation of the
coefficients of the one loop graviton energy. We begin with the coefficient
$a$ defined by%
\begin{equation}
a:=r_{t}\int_{r_{t}}^{+\infty}dr\frac{r^{2}}{\sqrt{1-\frac{b\left(  r\right)
}{r}}}\left[  U_{1}^{2}\left(  r\right)  +U_{2}^{2}\left(  r\right)  \right]
,
\end{equation}
which, with the help of Eqs. $\left(  \ref{potentials}\right)  $, becomes%
\begin{equation}
a=r_{t}\int_{r_{t}}^{+\infty}dr\frac{r^{2}}{\sqrt{1-\left(  \frac{r_{t}}%
{r}\right)  ^{1+\frac{1}{\omega}}}}\left[  2c_{1}^{2}\left(  r\right)
+2\left(  9+\frac{6}{\omega}+\frac{5}{\omega^{2}}\right)  c_{2}^{2}\left(
r\right)  +\frac{8}{\omega}c_{1}\left(  r\right)  c_{2}\left(  r\right)
\right]  ,
\end{equation}
where%
\begin{equation}
c_{1}\left(  r\right)  =\frac{6}{r^{2}}\left(  1-\left(  \frac{r_{t}}%
{r}\right)  ^{1+\frac{1}{\omega}}\right)  \qquad c_{2}\left(  r\right)
=\frac{1}{2r^{2}}\left(  \frac{r_{t}}{r}\right)  ^{1+\frac{1}{\omega}}.
\end{equation}
Defining the dimensionless variable $\rho=\frac{r}{r_{t}}$, we can write%
\[
a_{1}=r_{t}\int_{r_{t}}^{+\infty}dr\frac{r^{2}}{\sqrt{1-\left(  \frac{r_{t}%
}{r}\right)  ^{1+\frac{1}{\omega}}}}\left[  2c_{1}^{2}\left(  r\right)
\right]  =72\int_{1}^{+\infty}d\rho\left(  \rho^{\left(  1+\frac{1}{\omega
}\right)  }-1\right)  ^{\frac{3}{2}}\rho^{-2-\frac{3}{2}\left(  1+\frac
{1}{\omega}\right)  }%
\]%
\begin{equation}
=72\frac{\omega}{1+\omega}B\left(  \frac{5}{2},\frac{\omega}{1+\omega}\right)
.
\end{equation}%
\[
a_{2}=r_{t}\int_{r_{t}}^{+\infty}dr\frac{r^{2}}{\sqrt{1-\left(  \frac{r_{t}%
}{r}\right)  ^{1+\frac{1}{\omega}}}}\left[  2\left(  9+\frac{6}{\omega}%
+\frac{5}{\omega^{2}}\right)  c_{2}^{2}\left(  r\right)  \right]  =\frac{1}%
{2}\left(  9+\frac{6}{\omega}+\frac{5}{\omega^{2}}\right)  \int_{1}^{+\infty
}d\rho\left(  \rho^{\left(  1+\frac{1}{\omega}\right)  }-1\right)  ^{-\frac
{1}{2}}\rho^{-2-\frac{3}{2}\left(  1+\frac{1}{\omega}\right)  }%
\]%
\begin{equation}
\frac{1}{2}\left(  9+\frac{6}{\omega}+\frac{5}{\omega^{2}}\right)
\frac{\omega}{1+\omega}B\left(  \frac{1}{2},\frac{3\omega+2}{1+\omega}\right)
\end{equation}
and%
\[
a_{3}=r_{t}\int_{r_{t}}^{+\infty}dr\frac{r^{2}}{\sqrt{1-\left(  \frac{r_{t}%
}{r}\right)  ^{1+\frac{1}{\omega}}}}\left[  \frac{8}{\omega}c_{1}\left(
r\right)  c_{2}\left(  r\right)  \right]  =\frac{24}{\omega}\int_{1}^{+\infty
}d\rho\left(  \rho^{\left(  1+\frac{1}{\omega}\right)  }-1\right)  ^{\frac
{1}{2}}\rho^{-2-\frac{3}{2}\left(  1+\frac{1}{\omega}\right)  }%
\]%
\begin{equation}
=\frac{24}{1+\omega}B\left(  \frac{3}{2},\frac{2\omega+1}{1+\omega}\right)  ,
\end{equation}
so that%
\begin{equation}
a\equiv a\left(  \omega\right)  =\frac{\omega}{1+\omega}\left[  72B\left(
\frac{5}{2},\frac{\omega}{1+\omega}\right)  +\frac{1}{2}\left(  9+\frac
{6}{\omega}+\frac{5}{\omega^{2}}\right)  B\left(  \frac{1}{2},\frac{3\omega
+2}{1+\omega}\right)  +\frac{24}{\omega}B\left(  \frac{3}{2},\frac{2\omega
+1}{1+\omega}\right)  \right]  ,
\end{equation}
where $B\left(  x,y\right)  $ is the beta function$.$. The same procedure
applies to the coefficient $b$.%
\begin{equation}
b:=-2\int_{1}^{+\infty}d\rho\frac{\rho^{2}}{\sqrt{1-\left(  \frac{1}{\rho
}\right)  ^{1+\frac{1}{\omega}}}}\left[  U_{1}^{2}\left(  \rho\right)
\ln\left\vert U_{1}\left(  \rho\right)  \right\vert +U_{2}^{2}\left(
\rho\right)  \ln\left(  U_{2}\left(  \rho\right)  \right)  \right]  .
\end{equation}
We can separate the logarithmic functions%
\begin{equation}
\left\{
\begin{array}
[c]{c}%
\ln\left\vert U_{1}\left(  \rho\right)  \right\vert =\underset{b_{1a}%
}{\underbrace{\ln\left\vert 12\omega\left(  \rho^{\left(  1+\frac{1}{\omega
}\right)  }-1\right)  +\left(  1-3\omega\right)  \right\vert }}-\underset
{b_{1b}}{\underbrace{\ln\left(  2\omega\right)  }}-\underset{b_{1c}%
}{\underbrace{\left(  3+\frac{1}{\omega}\right)  \ln\rho}}\\
\\
\ln\left\vert U_{2}\left(  \rho\right)  \right\vert =\underset{b_{2a}%
}{\underbrace{\ln\left\vert 12\omega\left(  \rho^{\left(  1+\frac{1}{\omega
}\right)  }-1\right)  +3\left(  1+\omega\right)  \right\vert }}-\underset
{b_{2b}}{\underbrace{\ln\left(  2\omega\right)  }}-\underset{b_{2c}%
}{\underbrace{\left(  3+\frac{1}{\omega}\right)  \ln\rho}}%
\end{array}
\right.
\end{equation}
and compute separately the different terms. $b_{1b}+b_{2b}$ simply becomes%
\begin{equation}
b_{1b}+b_{2b}=2\ln\left(  2\omega\right)  \int_{1}^{+\infty}d\rho\frac
{\rho^{2}\left[  U_{1}^{2}\left(  \rho\right)  +U_{2}^{2}\left(  \rho\right)
\right]  }{\sqrt{1-\left(  \frac{1}{\rho}\right)  ^{1+\frac{1}{\omega}}}}%
=\ln\left(  2\omega\right)  a\left(  \omega\right)  .
\end{equation}
The coefficient $b_{1c}+b_{2c}$ is%
\begin{equation}
b_{1c}+b_{2c}=\left(  3+\frac{1}{\omega}\right)  \int_{1}^{+\infty}d\rho
\frac{\rho^{2}\left[  U_{1}^{2}\left(  \rho\right)  +U_{2}^{2}\left(
\rho\right)  \right]  }{\sqrt{1-\left(  \frac{1}{\rho}\right)  ^{1+\frac
{1}{\omega}}}}\ln\rho,
\end{equation}
namely%
\begin{equation}
\left(  3+\frac{1}{\omega}\right)  \int_{1}^{+\infty}d\rho\frac{\rho^{2}%
}{\sqrt{1-\left(  \frac{1}{\rho}\right)  ^{1+\frac{1}{\omega}}}}\left[
2c_{1}^{2}\left(  \rho\right)  +2\left(  9+\frac{6}{\omega}+\frac{5}%
{\omega^{2}}\right)  c_{2}^{2}\left(  \rho\right)  +\frac{8}{\omega}%
c_{1}\left(  \rho\right)  c_{2}\left(  \rho\right)  \right]  \ln\rho
\end{equation}
The first term, $b_{f,1}$ will be summed with the coefficient $a$, while the
second term can be written as $b_{f,b1}+b_{f,b2}+b_{f,b3}$ with%
\[
2\left(  3+\frac{1}{\omega}\right)  \int_{1}^{+\infty}d\rho\frac{\rho^{2}%
c_{1}^{2}\left(  \rho\right)  }{\sqrt{1-\left(  \frac{1}{\rho}\right)
^{1+\frac{1}{\omega}}}}\ln\rho=72\left(  3+\frac{1}{\omega}\right)  \left(
\frac{\omega}{1+\omega}\right)  ^{2}\int_{1}^{+\infty}d\rho\left(
\rho^{\left(  1+\frac{1}{\omega}\right)  }-1\right)  ^{\frac{3}{2}}%
\rho^{-2-\frac{3}{2}\left(  1+\frac{1}{\omega}\right)  }\ln\rho
\]%
\[
=72\left(  3+\frac{1}{\omega}\right)  \left(  \frac{\omega}{1+\omega}\right)
^{2}\int_{0}^{+\infty}dss^{\frac{3}{2}}\left(  1+s\right)  ^{-\frac{7\omega
+5}{2\left(  1+\omega\right)  }}\ln\left(  1+s\right)
\]%
\begin{equation}
=72\left(  3+\frac{1}{\omega}\right)  \left(  \frac{\omega}{1+\omega}\right)
^{2}B\left(  \frac{5}{2},\frac{\omega}{1+\omega}\right)  \left[  \Psi\left(
\frac{7\omega+5}{2\left(  1+\omega\right)  }\right)  -\Psi\left(  \frac
{\omega}{1+\omega}\right)  \right]  .
\end{equation}%
\[
\left(  3+\frac{1}{\omega}\right)  2\left(  9+\frac{6}{\omega}+\frac{5}%
{\omega^{2}}\right)  \int_{1}^{+\infty}d\rho\frac{\rho^{2}c_{2}^{2}\left(
\rho\right)  }{\sqrt{1-\left(  \frac{1}{\rho}\right)  ^{1+\frac{1}{\omega}}}%
}\ln\rho
\]%
\[
=\left(  3+\frac{1}{\omega}\right)  \frac{1}{2}\left(  9+\frac{6}{\omega
}+\frac{5}{\omega^{2}}\right)  \int_{1}^{+\infty}d\rho\left(  \rho^{\left(
1+\frac{1}{\omega}\right)  }-1\right)  ^{-\frac{1}{2}}\rho^{-2-\frac{3}%
{2}\left(  1+\frac{1}{\omega}\right)  }\ln\rho
\]%
\[
=\left(  3+\frac{1}{\omega}\right)  \frac{1}{2}\left(  9+\frac{6}{\omega
}+\frac{5}{\omega^{2}}\right)  \left(  \frac{\omega}{1+\omega}\right)
^{2}\int_{0}^{+\infty}dss^{-\frac{1}{2}}\left(  1+s\right)  ^{-\frac
{7\omega+5}{2\left(  1+\omega\right)  }}\ln\left(  1+s\right)
\]%
\begin{equation}
=\left(  3+\frac{1}{\omega}\right)  \frac{1}{2}\left(  9+\frac{6}{\omega
}+\frac{5}{\omega^{2}}\right)  \left(  \frac{\omega}{1+\omega}\right)
^{2}B\left(  \frac{1}{2},\frac{3\omega+2}{1+\omega}\right)  \left[
\Psi\left(  \frac{7\omega+5}{2\left(  1+\omega\right)  }\right)  -\Psi\left(
\frac{3\omega+2}{1+\omega}\right)  \right]
\end{equation}
and%
\[
\left(  3+\frac{1}{\omega}\right)  \frac{8}{\omega}\int_{1}^{+\infty}%
d\rho\frac{\rho^{2}c_{1}\left(  \rho\right)  c_{2}\left(  \rho\right)  }%
{\sqrt{1-\left(  \frac{1}{\rho}\right)  ^{1+\frac{1}{\omega}}}}\ln\rho=\left(
3+\frac{1}{\omega}\right)  \frac{24}{\omega}\int_{1}^{+\infty}d\rho\left(
\rho^{\left(  1+\frac{1}{\omega}\right)  }-1\right)  ^{\frac{1}{2}}%
\rho^{-2-\frac{3}{2}\left(  1+\frac{1}{\omega}\right)  }\ln\rho
\]%
\[
=\left(  3+\frac{1}{\omega}\right)  \frac{24}{\omega}\left(  \frac{\omega
}{1+\omega}\right)  ^{2}\int_{0}^{+\infty}dss^{\frac{1}{2}}\left(  1+s\right)
^{-\frac{7\omega+5}{2\left(  1+\omega\right)  }}\ln\left(  1+s\right)
\]%
\begin{equation}
=\left(  3+\frac{1}{\omega}\right)  \frac{24}{\omega}\left(  \frac{\omega
}{1+\omega}\right)  ^{2}B\left(  \frac{3}{2},\frac{2\omega+1}{1+\omega
}\right)  \left[  \Psi\left(  \frac{7\omega+5}{2\left(  1+\omega\right)
}\right)  -\Psi\left(  \frac{2\omega+1}{1+\omega}\right)  \right]  ,
\end{equation}
where we have used the following relation%
\begin{equation}
\int_{0}^{+\infty}dxx^{\mu-1}\frac{\ln\left(  \gamma+x\right)  }{\left(
\gamma+x\right)  ^{\nu}}=\gamma^{\mu-\nu}B\left(  \mu,\nu-\mu\right)  \left[
\Psi\left(  \nu\right)  -\Psi\left(  \nu-\mu\right)  +\ln\gamma\right]  ,
\end{equation}
where $\Psi$ is the digamma function. Globally, the coefficient $b_{1c}%
+b_{2c}$ becomes%
\[
=\left(  3+\frac{1}{\omega}\right)  \left(  \frac{\omega}{1+\omega}\right)
^{2}\left\{  72B\left(  \frac{5}{2},\frac{\omega}{1+\omega}\right)  \left[
\Psi\left(  \frac{7\omega+5}{2\left(  1+\omega\right)  }\right)  -\Psi\left(
\frac{\omega}{1+\omega}\right)  \right]  \right.
\]%
\begin{equation}
\left.  \frac{1}{2}\left(  9+\frac{6}{\omega}+\frac{5}{\omega^{2}}\right)
B\left(  \frac{1}{2},\frac{3\omega+2}{1+\omega}\right)  \left[  \Psi\left(
\frac{7\omega+5}{2\left(  1+\omega\right)  }\right)  -\Psi\left(
\frac{3\omega+2}{1+\omega}\right)  \right]  +\frac{24}{\omega}B\left(
\frac{3}{2},\frac{2\omega+1}{1+\omega}\right)  \left[  \Psi\left(
\frac{7\omega+5}{2\left(  1+\omega\right)  }\right)  -\Psi\left(
\frac{2\omega+1}{1+\omega}\right)  \right]  \right\}  .
\end{equation}
The term $b_{1a}+b_{2b}$ is more complicated. The logarithmic term is of the
form $\ln\left(  \gamma_{1,2}+\delta x\right)  $. It is useful to re-express
it in terms of hypergeometric function, namely%
\begin{equation}
\ln\left(  \gamma_{1,2}+\delta x\right)  =\delta\frac{x}{\gamma_{1,2}}%
\ \,_{2}F_{1}\ \left(  1,1;2;-\delta\frac{x}{\gamma_{1,2}}\right)  +\ln
\gamma_{1,2},
\end{equation}
with $\delta=12\omega$ and%
\begin{equation}
\left\{
\begin{array}
[c]{c}%
\gamma_{1}=1-3\omega\\
\\
\gamma_{2}=3\left(  1+\omega\right)
\end{array}
\right.  .
\end{equation}
Expression%
\begin{equation}
\ln\gamma_{1}\int_{1}^{+\infty}d\rho\frac{\rho^{2}}{\sqrt{1-\left(  \frac
{1}{\rho}\right)  ^{1+\frac{1}{\omega}}}}\left[  c_{1}^{2}\left(  \rho\right)
+\left(  \frac{1}{\omega}-3\right)  ^{2}c_{2}^{2}\left(  \rho\right)
+2\left(  \frac{1}{\omega}-3\right)  c_{1}\left(  \rho\right)  c_{2}\left(
\rho\right)  \right]
\end{equation}
becomes%
\begin{equation}
\ln\gamma_{1}\frac{\omega}{1+\omega}\left[  36B\left(  \frac{5}{2}%
,\frac{\omega}{1+\omega}\right)  +\left(  \frac{1}{\omega}-3\right)
^{2}B\left(  \frac{1}{2},\frac{3\omega+2}{1+\omega}\right)  +6\left(  \frac
{1}{\omega}-3\right)  B\left(  \frac{3}{2},\frac{2\omega+1}{1+\omega}\right)
\right]
\end{equation}
and expression%
\begin{equation}
\ln\gamma_{2}\int_{1}^{+\infty}d\rho\frac{\rho^{2}}{\sqrt{1-\left(  \frac
{1}{\rho}\right)  ^{1+\frac{1}{\omega}}}}\left[  c_{1}^{2}\left(  \rho\right)
+\left(  \frac{3}{\omega}+3\right)  ^{2}c_{2}^{2}\left(  \rho\right)
+2\left(  \frac{3}{\omega}+3\right)  c_{1}\left(  \rho\right)  c_{2}\left(
\rho\right)  \right]
\end{equation}
becomes%
\begin{equation}
\ln\gamma_{2}\frac{\omega}{1+\omega}\left[  36B\left(  \frac{5}{2}%
,\frac{\omega}{1+\omega}\right)  +\left(  \frac{3}{\omega}+3\right)
^{2}B\left(  \frac{1}{2},\frac{3\omega+2}{1+\omega}\right)  +6\left(  \frac
{3}{\omega}+3\right)  B\left(  \frac{3}{2},\frac{2\omega+1}{1+\omega}\right)
\right]  .
\end{equation}
Concerning the hypergeometric part, after having defined the variable
$s=\rho^{\left(  1+\frac{1}{\omega}\right)  }-1$, we use the following formula%
\[
\int_{0}^{+\infty}dx\frac{x^{\alpha-1}}{\left(  x+z\right)  ^{\rho}}%
\ \,_{2}F_{1}\left(  a,b;c;-\omega x\right)  dx=z^{\alpha-\rho}B\left(
\alpha,\rho-\alpha\right)  \ \,_{3}F_{2}\ \left(  a,b,\alpha;c,\alpha
-\rho+1;\omega z\right)
\]%
\begin{equation}
+\omega^{\rho-\alpha}\frac{\Gamma\left(  c\right)  \Gamma\left(  a-\alpha
+\rho\right)  \Gamma\left(  b-\alpha+\rho\right)  \Gamma\left(  \alpha
-\rho\right)  }{\Gamma\left(  a\right)  \Gamma\left(  b\right)  \Gamma\left(
c-\alpha+\rho\right)  }\ \,_{3}F_{2}\ \left(  a-\alpha+\rho,b-\alpha+\rho
,\rho;c-\alpha+\rho,\rho-\alpha+1;\omega z\right)  ,
\end{equation}
which, in our specific case, becomes%
\[
\frac{\delta}{\gamma_{1,2}}\int_{0}^{+\infty}ds\frac{s^{\alpha-1}}{\left(
s+1\right)  ^{\rho}}s\,_{2}F_{1}\left(  1,1;2;-\delta\frac{s}{\gamma_{1,2}%
}\right)  ds=B\left(  \alpha,\rho-\alpha\right)  \ \,_{3}F_{2}\ \left(
1,1,\alpha;2,\alpha-\rho+1;\frac{\delta}{\gamma_{1,2}}\right)
\]%
\begin{equation}
+\left(  \frac{\delta}{\gamma_{1,2}}\right)  ^{\rho-\alpha}\frac{\Gamma\left(
1-\alpha+\rho\right)  \Gamma\left(  1-\alpha+\rho\right)  \Gamma\left(
\alpha-\rho\right)  }{\Gamma\left(  2-\alpha+\rho\right)  }\ \,_{3}%
F_{2}\ \left(  1-\alpha+\rho,1-\alpha+\rho,\rho;2-\alpha+\rho,1-\alpha
+\rho;\frac{\delta}{\gamma_{1,2}}\right)  .
\end{equation}
Therefore for $\gamma_{1}$ we get%
\begin{equation}
\frac{\delta}{\gamma_{1}}\int_{1}^{+\infty}d\rho\frac{\,_{2}F_{1}\left(
1,1;2;-\frac{\delta}{\gamma_{1}}\left(  \rho^{\left(  1+\frac{1}{\omega
}\right)  }-1\right)  \right)  \rho^{2}}{\sqrt{1-\left(  \frac{1}{\rho
}\right)  ^{1+\frac{1}{\omega}}}}\left(  \rho^{\left(  1+\frac{1}{\omega
}\right)  }-1\right)  \left[  c_{1}^{2}\left(  \rho\right)  +\left(  \frac
{1}{\omega}-3\right)  ^{2}c_{2}^{2}\left(  \rho\right)  +2\left(  \frac
{1}{\omega}-3\right)  c_{1}\left(  \rho\right)  c_{2}\left(  \rho\right)
\right]  .
\end{equation}
Every piece leads to: $a)$%
\[
\frac{\delta}{\gamma_{1}}\int_{1}^{+\infty}d\rho\frac{\rho^{2}c_{1}^{2}\left(
\rho\right)  }{\sqrt{1-\left(  \frac{1}{\rho}\right)  ^{1+\frac{1}{\omega}}}%
}\left(  \rho^{\left(  1+\frac{1}{\omega}\right)  }-1\right)  \,_{2}%
F_{1}\left(  1,1;2;-\frac{\delta}{\gamma_{1}}\left(  \rho^{\left(  1+\frac
{1}{\omega}\right)  }-1\right)  \right)
\]%
\[
=36\frac{\delta}{\gamma_{1}}\left(  \frac{\omega}{1+\omega}\right)  \int
_{0}^{+\infty}dss^{\frac{5}{2}}\left(  1+s\right)  ^{-\frac{7\omega
+5}{2\left(  1+\omega\right)  }}\,_{2}F_{1}\left(  1,1;2;-\frac{\delta}%
{\gamma_{1}}s\right)
\]%
\[
=36\frac{\delta}{\gamma_{1}}\left(  \frac{\omega}{1+\omega}\right)  \left[
B\left(  \frac{7}{2},-\frac{2}{1+\omega}\right)  \ \,_{3}F_{2}\ \left(
1,1,\frac{7}{2};2,\frac{\omega+3}{1+\omega};\frac{\delta}{\gamma_{1}}\right)
\right.
\]%
\begin{equation}
\left.  +\left(  \frac{\delta}{\gamma_{1}}\right)  ^{-\frac{2}{1+\omega}}%
\frac{\left[  \Gamma\left(  \frac{\omega-1}{1+\omega}\right)  \right]
^{2}\Gamma\left(  \frac{2}{1+\omega}\right)  }{\Gamma\left(  \frac{2\omega
}{1+\omega}\right)  }\ \,_{3}F_{2}\ \left(  \frac{\omega-1}{1+\omega}%
,\frac{\omega-1}{1+\omega},\frac{7\omega+5}{2\left(  1+\omega\right)  }%
;\frac{2\omega}{1+\omega},\frac{\omega-1}{1+\omega};\frac{\delta}{\gamma_{1}%
}\right)  \right]  , \label{delta1}%
\end{equation}
$b)$%
\[
\frac{\delta}{\gamma_{1}}\left(  \frac{1}{\omega}-3\right)  ^{2}\int
_{1}^{+\infty}d\rho\frac{\rho^{2}c_{2}^{2}\left(  \rho\right)  }%
{\sqrt{1-\left(  \frac{1}{\rho}\right)  ^{1+\frac{1}{\omega}}}}\left(
\rho^{\left(  1+\frac{1}{\omega}\right)  }-1\right)  \,_{2}F_{1}\left(
1,1;2;-\frac{\delta}{\gamma_{1}}\left(  \rho^{\left(  1+\frac{1}{\omega
}\right)  }-1\right)  \right)
\]%
\[
=\frac{\delta\gamma_{1}}{4\omega\left(  1+\omega\right)  }\int_{0}^{+\infty
}dss^{\frac{1}{2}}\left(  1+s\right)  ^{-\frac{7\omega+5}{2\left(
1+\omega\right)  }}\,_{2}F_{1}\left(  1,1;2;-\frac{\delta}{\gamma_{1}%
}s\right)  =\frac{\delta\gamma_{1}}{4\omega\left(  1+\omega\right)  }\left[
B\left(  \frac{3}{2},\frac{2\omega+1}{1+\omega}\right)  \ \,_{3}F_{2}\ \left(
1,1,\frac{3}{2};2,-\frac{\omega}{1+\omega};\frac{\delta}{\gamma_{1}}\right)
\right.
\]%
\begin{equation}
\left.  +\left(  \frac{\delta}{\gamma_{1}}\right)  ^{-\frac{2\omega
+1}{1+\omega}}\frac{\left[  \Gamma\left(  \frac{3\omega+2}{1+\omega}\right)
\right]  ^{2}\Gamma\left(  -\frac{2\omega+1}{1+\omega}\right)  }{\Gamma\left(
\frac{4\omega+3}{1+\omega}\right)  }\ \,_{3}F_{2}\ \left(  \frac{3\omega
+2}{1+\omega},\frac{3\omega+2}{1+\omega},\frac{7\omega+5}{2\left(
1+\omega\right)  };\frac{4\omega+3}{1+\omega},\frac{3\omega+2}{1+\omega}%
;\frac{\delta}{\gamma_{1}}\right)  \right]  \label{delta2}%
\end{equation}
and $c)$%
\[
\frac{\delta}{\gamma_{1}}\left(  \frac{1}{\omega}-3\right)  \int_{1}^{+\infty
}d\rho\frac{\rho^{2}c_{1}\left(  \rho\right)  c_{2}\left(  \rho\right)
}{\sqrt{1-\left(  \frac{1}{\rho}\right)  ^{1+\frac{1}{\omega}}}}\left(
\rho^{\left(  1+\frac{1}{\omega}\right)  }-1\right)  \,_{2}F_{1}\left(
1,1;2;-\frac{\delta}{\gamma_{1}}\left(  \rho^{\left(  1+\frac{1}{\omega
}\right)  }-1\right)  \right)
\]%
\[
=\frac{6\delta}{1+\omega}\int_{0}^{+\infty}dss^{\frac{3}{2}}\left(
1+s\right)  ^{-\frac{7\omega+5}{2\left(  1+\omega\right)  }}\,_{2}F_{1}\left(
1,1;2;-\frac{\delta}{\gamma_{1}}s\right)  =\frac{6\delta}{1+\omega}\left[
B\left(  \frac{5}{2},\frac{\omega}{1+\omega}\right)  \ \,_{3}F_{2}\ \left(
1,1,\frac{5}{2};2,\frac{1}{1+\omega};\frac{\delta}{\gamma_{1}}\right)
\right.
\]%
\begin{equation}
\left.  +\left(  \frac{\delta}{\gamma_{1}}\right)  ^{\frac{\omega}{1+\omega}%
}\frac{\left[  \Gamma\left(  \frac{2\omega+1}{1+\omega}\right)  \right]
^{2}\Gamma\left(  -\frac{\omega}{1+\omega}\right)  }{\Gamma\left(
\frac{3\omega+2}{1+\omega}\right)  }\ \,_{3}F_{2}\ \left(  \frac{2\omega
+1}{1+\omega},\frac{2\omega+1}{1+\omega},\frac{7\omega+5}{2\left(
1+\omega\right)  };\frac{3\omega+2}{1+\omega},\frac{2\omega+1}{1+\omega}%
;\frac{\delta}{\gamma_{1}}\right)  \right]  . \label{delta3}%
\end{equation}
The same set of expressions works for $\gamma_{2}$ too.

\section{The zeta function regularization}

\label{app2}In this appendix, we report details on computation leading to
expression $\left(  \ref{zeta}\right)  $. We begin with the following integral%
\begin{equation}
\rho\left(  \varepsilon\right)  =\left\{
\begin{array}
[c]{c}%
I_{+}=\mu^{2\varepsilon}\int_{0}^{+\infty}d\omega\frac{\omega^{2}}{\left(
\omega^{2}+U\left(  x\right)  \right)  ^{\varepsilon-\frac{1}{2}}}\\
\\
I_{-}=\mu^{2\varepsilon}\int_{0}^{+\infty}d\omega\frac{\omega^{2}}{\left(
\omega^{2}-U\left(  x\right)  \right)  ^{\varepsilon-\frac{1}{2}}}%
\end{array}
\right.  , \label{rho}%
\end{equation}
with $U\left(  x\right)  >0$.

\subsection{$I_{+}$ computation}

\label{app2a}If we define $t=\omega/\sqrt{U\left(  x\right)  }$, the integral
$I_{+}$ in Eq.$\left(  \ref{rho}\right)  $ becomes%
\[
\rho\left(  \varepsilon\right)  =\mu^{2\varepsilon}U\left(  x\right)
^{2-\varepsilon}\int_{0}^{+\infty}dt\frac{t^{2}}{\left(  t^{2}+1\right)
^{\varepsilon-\frac{1}{2}}}=\frac{1}{2}\mu^{2\varepsilon}U\left(  x\right)
^{2-\varepsilon}B\left(  \frac{3}{2},\varepsilon-2\right)
\]%
\begin{equation}
\frac{1}{2}\mu^{2\varepsilon}U\left(  x\right)  ^{2-\varepsilon}\frac
{\Gamma\left(  \frac{3}{2}\right)  \Gamma\left(  \varepsilon-2\right)
}{\Gamma\left(  \varepsilon-\frac{1}{2}\right)  }=\frac{\sqrt{\pi}}{4}U\left(
x\right)  ^{2}\left(  \frac{\mu^{2}}{U\left(  x\right)  }\right)
^{\varepsilon}\frac{\Gamma\left(  \varepsilon-2\right)  }{\Gamma\left(
\varepsilon-\frac{1}{2}\right)  },
\end{equation}
where we have used the following identities involving the beta function%
\begin{equation}
B\left(  x,y\right)  =2\int_{0}^{+\infty}dt\frac{t^{2x-1}}{\left(
t^{2}+1\right)  ^{x+y}}\qquad\operatorname{Re}x>0,\operatorname{Re}y>0
\end{equation}
related to the gamma function by means of%
\begin{equation}
B\left(  x,y\right)  =\frac{\Gamma\left(  x\right)  \Gamma\left(  y\right)
}{\Gamma\left(  x+y\right)  }.
\end{equation}
Taking into account the following relations for the $\Gamma$-function%
\begin{equation}
\Gamma\left(  \varepsilon-2\right)  =\frac{\Gamma\left(  1+\varepsilon\right)
}{\varepsilon\left(  \varepsilon-1\right)  \left(  \varepsilon-2\right)
},\qquad\Gamma\left(  \varepsilon-\frac{1}{2}\right)  =\frac{\Gamma\left(
\varepsilon+\frac{1}{2}\right)  }{\varepsilon-\frac{1}{2}}, \label{gamma}%
\end{equation}
and the expansion for small $\varepsilon$%
\[
\Gamma\left(  1+\varepsilon\right)  =1-\gamma\varepsilon+O\left(
\varepsilon^{2}\right)  ,\qquad\Gamma\left(  \varepsilon+\frac{1}{2}\right)
=\Gamma\left(  \frac{1}{2}\right)  -\varepsilon\Gamma\left(  \frac{1}%
{2}\right)  \left(  \gamma+2\ln2\right)  +O\left(  \varepsilon^{2}\right)
\]%
\begin{equation}
x^{\varepsilon}=1+\varepsilon\ln x+O\left(  \varepsilon^{2}\right)  ,
\end{equation}
where $\gamma$ is the Euler's constant, we find%
\begin{equation}
\rho\left(  \varepsilon\right)  =-\frac{U^{2}\left(  x\right)  }{16}\left[
\frac{1}{\varepsilon}+\ln\left(  \frac{\mu^{2}}{U\left(  x\right)  }\right)
+2\ln2-\frac{1}{2}\right]  .
\end{equation}

\subsection{$I_{-}$ computation}

\label{app2b}If we define $t=\omega/\sqrt{U\left(  x\right)  }$, the integral
$I_{-}$ in Eq.$\left(  \ref{rho}\right)  $ becomes%
\[
\rho\left(  \varepsilon\right)  =\mu^{2\varepsilon}U\left(  x\right)
^{2-\varepsilon}\int_{0}^{+\infty}dt\frac{t^{2}}{\left(  t^{2}-1\right)
^{\varepsilon-\frac{1}{2}}}=\frac{1}{2}\mu^{2\varepsilon}U\left(  x\right)
^{2-\varepsilon}B\left(  \varepsilon-2,\frac{3}{2}-\varepsilon\right)
\]%
\begin{equation}
\frac{1}{2}\mu^{2\varepsilon}U\left(  x\right)  ^{2-\varepsilon}\frac
{\Gamma\left(  \frac{3}{2}-\varepsilon\right)  \Gamma\left(  \varepsilon
-2\right)  }{\Gamma\left(  -\frac{1}{2}\right)  }=-\frac{1}{4\sqrt{\pi}%
}U\left(  x\right)  ^{2}\left(  \frac{\mu^{2}}{U\left(  x\right)  }\right)
^{\varepsilon}\Gamma\left(  \frac{3}{2}-\varepsilon\right)  \Gamma\left(
\varepsilon-2\right)  ,
\end{equation}
where we have used the following identity involving the beta function%
\begin{equation}
\frac{1}{p}B\left(  1-\nu-\frac{\mu}{p},\nu\right)  =\int_{1}^{+\infty
}dtt^{\mu-1}\left(  t^{p}-1\right)  ^{\nu-1}\qquad p>0,\operatorname{Re}%
\nu>0,\operatorname{Re}\mu<p-p\operatorname{Re}\nu
\end{equation}
and the reflection formula%
\begin{equation}
\Gamma\left(  z\right)  \Gamma\left(  1-z\right)  =-z\Gamma\left(  -z\right)
\Gamma\left(  z\right)
\end{equation}
From the first of Eqs.$\left(  \ref{gamma}\right)  $ and from the expansion
for small $\varepsilon$%
\[
\Gamma\left(  \frac{3}{2}-\varepsilon\right)  =\Gamma\left(  \frac{3}%
{2}\right)  \left(  1-\varepsilon\left(  -\gamma-2\ln2+2\right)  \right)
+O\left(  \varepsilon^{2}\right)
\]%
\begin{equation}
x^{\varepsilon}=1+\varepsilon\ln x+O\left(  \varepsilon^{2}\right)  ,
\end{equation}
we find%
\begin{equation}
\rho\left(  \varepsilon\right)  =-\frac{U^{2}\left(  x\right)  }{16}\left[
\frac{1}{\varepsilon}+\ln\left(  \frac{\mu^{2}}{U\left(  x\right)  }\right)
+2\ln2-\frac{1}{2}\right]  .
\end{equation}


\begin{thebibliography}{99}                                                                                               %


\bibitem {expansion}A. G. Riess \textit{et al.}, \textsl{Astron. J.
}\textbf{116}, 1009 (1998);

\bibitem {MT}M. S. Morris and K. S. Thorne, \textsl{Am. J. Phys.} \textbf{56},
395 (1988).

\bibitem {MTY}M. S. Morris, K. S. Thorne, and U. Yurtsever, \textsl{Phys. Rev.
Lett.} \textbf{61}, 1446 (1988).

\bibitem {Lobo}F. S. N. Lobo, \textsl{Phys. Rev.} \textbf{D} \textbf{71
}124022, (2005); gr-qc/0506001. F. S. N. Lobo, \textsl{Phys. Rev.} \textbf{D}
\textbf{71 }084011, (2005); gr-qc/0502099.

\bibitem {Kuhfittig}P. K. F. Kuhfittig, \textsl{Class.Quant.Grav. }\textbf{23}
5853, (2006); gr-qc/0608055.

\bibitem {Sushkov}S. Sushkov, \textsl{Phys. Rev.} \textbf{D} \textbf{71
}043520, (2005), gr-qc/0502084.

\bibitem {Remo0}R. Garattini, \textsl{Class.Quant.Grav. }\textbf{22} 1105,
(2005); gr-qc/0501105.

\bibitem {RemoFrancisco}R. Garattini and F. Lobo, \textit{Self sustained phantom wormholes in semi-classical gravity} (to be submitted).

\bibitem {FOP}C. J. Fewster, K. D. Olum and M. J. Pfenning, \textit{Averaged
null energy condition in spacetimes with boundaries}. gr-qc/0609007.

\bibitem {HPS}D. Hochberg, A. Popov and S. V. Sushkov, \textsl{Phys. Rev.
Lett.} \textbf{78 }(1997) 2050, gr-qc/9701064.

\bibitem {KS}N. R. Khusnutdinov and S. V. Sushkov, \textsl{Phys. Rev.}
\textbf{D} \textbf{65 }084028, (2002), hep-th/0202068.

\bibitem {MVisser}M. Visser, \textit{Lorentzian Wormholes} (AIP Press, New
York, 1995) 64.

\bibitem {AndersonBrill}P. R. Anderson and D. R. Brill, \textsl{Phys.Rev.} D
\textbf{56} (1997) 4824, gr-qc/9610074 .

\bibitem {ADM}R. Arnowitt, S. Deser, and C. W. Misner, in \textit{Gravitation:
An Introduction to Current Research} edited by L. Witten (John Wiley \& Sons,
Inc., New York, 1962).

\bibitem {Remo}R. Garattini,\textsl{ Phys.Rev.} \textbf{D} \textbf{59} 104019,
(1999), hep-th/9902006.

\bibitem {Remo1}R. Garattini, \textsl{Int. J. Mod. Phys.} \textbf{A}
\textbf{18} (1999) 2905, gr-qc/9805096.

\bibitem {BergerEbin}M. Berger and D. Ebin, \textsl{J. Diff. Geom.}
\textbf{3}, 379 (1969).

\bibitem {York}J. W. York Jr., \textsl{J. Math. Phys.}, \textbf{14}, 4 (1973),
\textsl{Ann. Inst. Henri Poincar\'{e}} \textbf{A} \textbf{21} (1974) 319.

\bibitem {RW}T. Regge and J. A. Wheeler, \textsl{Phys. Rev.} \textbf{108},
1063 (1957).

\bibitem {BMM}M. Bordag, U. Mohideen and V.M. Mostepanenko, \textsl{Phys.
Rep.} \textbf{353}, 1 (2001).

\bibitem {Cherednikov}I.O. Cherednikov, \textsl{Acta Physica Slovaca},
\textbf{52}, (2002), 221.

\bibitem {HawEll}S.W. Hawking and G.F.R. Ellis, \textit{The Large Scale
Structure of Spacetime} (Cambridge Univ. Press, 1973).

\bibitem {GR}I.S. Gradshteyn and I.M. Ryzhik, \textit{Table of Integrals,
Series, and Products }(corrected and enlarged edition), edited by A. Jeffrey
(Academic Press, Inc.).

\bibitem {PBM}A. P. Prudnikov, Yu. A. Brychkov and O.I. Marichev,
\textit{Integrals and Series,Vol. 3: More Special Functions}, edited by Gordon
and Breach Science Publishers, Second Printing 1998.
\end{thebibliography}
\end{document}